\documentclass[11pt,fleqn]{article}

\usepackage{geometry}
\usepackage{a4wide}
\usepackage{amsmath}
\usepackage{amssymb}
\usepackage{amsthm}
\usepackage[round]{natbib}
\usepackage{rotating}
\usepackage{xcolor}
\usepackage{enumerate}
\usepackage{setspace}
\usepackage{bm}
\usepackage{bbm}
\usepackage{graphicx}
\usepackage{multicol} 
\usepackage[bottom]{footmisc}
\usepackage[tableposition=top]{caption}
\usepackage{subfigure}
\usepackage{multirow}
\usepackage{float}
\usepackage{hyperref}
\newcommand{\bA}{\bm A}

\newcommand{\bB}{\bm B}

\newcommand{\bC}{\bm C}

\newcommand{\bof}{\bm f}

\newcommand{\bI}{\bm I}

\newcommand{\bQ}{\bm Q}

\newcommand{\bR}{\bm R}

\newcommand{\bs}{\bm s}

\newcommand{\bu}{\bm u}

\newcommand{\bw}{\bm w}
\newcommand{\bX}{\bm X}
\newcommand{\bx}{\bm x}

\newcommand{\bZ}{\bm Z}
\newcommand{\bz}{\bm z}

\newcommand{\bdelta}{\bm \delta}

\newcommand{\bvarepsilon}{\bm \varepsilon}
\newcommand{\bzeta}{\bm \zeta}

\newcommand{\blambda}{\bm \lambda}
\newcommand{\bmu}{\bm \mu}

\newcommand{\bpi}{\bm \pi}

\newcommand{\btau}{\bm \tau}

\newcommand{\bGamma}{\bm \varGamma}

\newcommand{\bLambda}{\bm \varLambda}

\newcommand{\bSigma}{\bm \varSigma}

\newcommand{\bPhi}{\bm \varPhi}

\newcommand{\bOmega}{\bm \varOmega}

\newcommand{\abs}[1]{\left\lvert#1\right\rvert}
\newcommand{\norm}[1]{\left\lVert#1\right\rVert}

\newcommand{\E}{\mathbb{E}}

\newcommand{\citepos}[1]{\citeauthor{#1}'s \citeyearpar{#1}}

\newtheoremstyle{rem}%
{10pt}%
{10pt}%
{}%
{}%
{\bf}%
{.}%
{.5em}%
{}%

\theoremstyle{rem}
\newtheorem{remark}{Remark}

\title{High-Dimensional Forecasting in the Presence of \\Unit Roots and Cointegration}
\author{Stephan Smeekes\thanks{Department of Quantitative Economics, Maastricht University, P.O. Box 616, 6200 MD Maastricht, The Netherlands. E-mail: \href{mailto:s.smeekes@maastrichtuniversity.nl}{\textcolor{blue}{s.smeekes@maastrichtuniversity.nl}}}
\and Etienne Wijler\thanks{Department of Quantitative Economics, Maastricht University, P.O. Box 616, 6200 MD Maastricht, The Netherlands. E-mail: \href{mailto:e.wijler@maastrichtuniversity.nl}{\textcolor{blue}{e.wijler@maastrichtuniversity.nl}}}}
\date{Department of Quantitative Economics\\
Maastricht University}

\begin{document}

\maketitle

\begin{abstract}
We investigate how the possible presence of unit roots and cointegration affects forecasting with Big Data. As most macroeoconomic time series are very persistent and may contain unit roots, a proper handling of unit roots and cointegration is of paramount importance for macroeconomic forecasting. The high-dimensional nature of Big Data complicates the analysis of unit roots and cointegration in two ways. First, transformations to stationarity require performing many unit root tests, increasing room for errors in the classification. Second, modelling unit roots and cointegration directly is more difficult, as standard high-dimensional techniques such as factor models and penalized regression are not directly applicable to (co)integrated data and need to be adapted. We provide an overview of both issues and review methods proposed to address these issues. These methods are also illustrated with two empirical applications.

\end{abstract}

\section{Introduction} \label{sec:intro}
We investigate forecasting with Big Data when the series in the dataset may contain unit roots and be cointegrated. As most macroeoconomic time series are at least very persistent, and may contain unit roots, a proper handling of unit roots and cointegration is of paramount importance in macroeconomic forecasting. The theory of unit roots and cointegration in small systems is well-developed and numerous reference works exist to guide the practitioner, see for example \citet{Enders08} or \citet{Hamilton94} for comprehensive treatments.

We discuss the problems that arise when extending the analysis to high-dimensional data and consider solutions that have been proposed in the literature. In particular, we discuss the applicability of the proposed methods for macroeconomic forecasting, reviewing relevant theoretical properties and practical issues. Moreover, by considering two big data applications ---that are very different in spirit--- we illustrate the issues and analyze the performance of the various methods in practically relevant situations.

The empirical literature dealing with unit roots and cointegration can essentially be split into two different philosophies. The first approach is to apply an appropriate transformation to each series such that one can work with stationary time series, with the most common transformation taking first differences of a series with a unit root. This is the most common approach in high-dimensional forecasting, as it only involves ``straightforward'' unit root or stationarity testing on each series. Indeed, commonly used Big Data such as the FRED-MD and -QD datasets \citep{McCrackenNg16} already come with pre-determined transformation codes to achieve stationarity. While this approach appears to be conceptually simple, we will argue that there are apparently minor issues that are often ignored in practice, but which can have a big impact on the performance of consequent forecasts, in particular when working with less established datasets.

The second approach is to model unit root and cointegration properties directly. In small systems, this is commonly done through vector error correction models (VECM), often using the popular maximum likelihood methodology developed by \citet{Johansen95}. The rationale for this seemingly more complicated approach is that ignoring long-run relations between the variables, as is done in the first approach, means not incorporating all information into the forecaster's model, which may have a detrimental effect on the forecast quality. Extending these techniques for modelling cointegration to high-dimensional settings requires a careful rethink of how cointegration can be viewed in high dimensions, and is an ongoing area of research. We will discuss recent contributions in this area and analyze the respective merits and drawback of each method.

While the importance of the concept of cointegration for macroeconometric analysis cannot be understated, one might argue that for the specific goal of forecasting it is not crucial. In the low-dimensional time series literature a large body of literature exists which compares the relative merits of the two philosophical approaches for forecasting, see for instance \citet{ClementsHendry95}, \citet{ChristoffersenDiebold98}, \citet{DieboldKilian00} and the references therein. Generally, the conclusion is mixed, with the performance of each approach varying depending on forecast horizon, dimensions of the models, estimation accuracy, and even specific applications and datasets. As this is no different in a high-dimensional context, we make no attempt to classify one of these approaches as superior. Instead, we aim to provide the practitioner with an overview of tools available to follow either line of thought.

One could discern a third approach to unit roots and cointegration, which is to ignore unit roots all together and estimate all forecasting models in levels. While this approach is at first glance close to the first approach and one might have valid reasons to prefer this approach, we do not recommend this in high-dimensional problems. If cointegration is not present in (parts of) the data, these methods may be very sensitive to spurious regression. The higher the dimensions of the data, the more likely that spurious regression becomes an issue. In particular, given that many methods discussed in this book perform some sort of dimensionality reduction or variable selection, this may actually increase the likelihood of obtaining spurious results. For instance, \citet{SmeekesWijler18} investigate the sensitivity of penalized regression methods to spurious results, and find that their variable selection mechanisms cannot properly distinguish between cointegrated and spurious regressors. Low-dimensional solutions such as always including lagged levels to avoid spurious regression are not possible in high-dimensional systems, as it would require including too many variables, and the applied dimensionality reduction or variable selection techniques might not be able to retain the lagged levels in the model. As such, we do not consider the approach of estimating everything in levels further.
\footnote{Obviously, this caveat does not mean that forecasting in levels does not yield good results for specific applications. The applied researcher is free to apply any of the methods discussed in this book directly to (suspected) unit root series, but should simply be wary of the results.}

We also illustrate the discussed methods by two empirical applications. In the first we forecast several U.S.~macroeconomic variables using the FRED-MD database. This application tests the methods in a known macroeconomic context, thus serving as a benchmark. In our second application, we consider nowcasting unemployment using a dataset constructed from Google Trends with frequencies of unemployment-related search terms. This second application not only serves to highlight the potential of ``modern'' Big Data sources for macroeconomic forecasting, but also illustrates that in such Big Data applications, we have little theoretical guidance to decide on unit root and cointegration properties, and proper data-driven methods are needed.

Note that, as is common in the related high-dimensional literature, we focus explicitly on point forecasts. As distributional theory changes when unit roots are present, performing interval forecasts in the presence of unit roots and cointegration is a much more challenging -- and largely unresolved -- issue in the high-dimensional setting, especially as it adds to the complications of performing inference in high dimensions already present without unit roots. Given the sparsity of literature on this topic, we do not consider interval prediction. This is clearly a very important avenue for future research.

The remainder is organized as follows. Section \ref{sec:setup} describes the general setup and introduces the cointegration model, along with some useful representations for later use. We discuss how to transform high-dimensional datasets to stationarity in Section \ref{sec:trans}, while Section \ref{sec:coint} introduces high-dimensional approaches for modelling cointegration. In Section \ref{sec:appl} we apply the discussed methods to our two empirical forecasting exercises. Finally, Section \ref{sec:conc} concludes.

\section{General Setup} \label{sec:setup}
In this section we describe a general model for cointegration to be used throughout. Next to defining the model in the classical error correction form, we also consider alternative representations that will be useful later. As is common in the literature, we denote a time series as $I(d)$ if it has to be differenced $d$ times to achieve stationarity, and we will use $I(1)$ interchangeably with a unit root process, and $I(0)$ with a stationary process.\footnote{In fact, $I(0)$ processes can be non-stationary for example through having time-varying unconditional variance. For ease of explanation we still use `stationary' to describe $I(0)$ processes though.}

Let $\bz_t$ denote an $N$-dimensional time series observed at time $t = 1, \ldots, T$. Assume that we can represent the series as
\begin{equation}\label{eq:model1}
\bz_t = \bmu + \btau t + \bzeta_t,
\end{equation}
where $\bmu$ is an $N$-dimensional vector of intercepts, $\btau$ is an $n$-dimensional vector of trend slopes, and $\bzeta_t$ is the $N$-dimensional purely stochastic time series. This stochastic component given is by
\begin{equation}\label{eq:model2}
\begin{split}
\Delta\bzeta_t &= \bA \bB^\prime \bzeta_{t-1} + \sum_{j=1}^p \bPhi_j \Delta \bzeta_{t-j} + \bvarepsilon_t,
\end{split}
\end{equation}
where $\bvarepsilon_t$ is the $N$-dimensional innovation vector. Generally the innovations $\bvarepsilon_t$ will be a martingale difference sequence, although we abstract from making too specific assumptions at this point. 

We can obtain the classical vector error correction model (VECM) for $\bz_t$ by substituting \eqref{eq:model1} into \eqref{eq:model2}:
\begin{equation}\label{eq:VECM}
\begin{split}
\Delta \bz_t &= \bA\bB^\prime \left(\bz_{t-1}-\bmu-\btau (t-1)\right) + \btau^* + \sum_{j=1}^p \bPhi_j \Delta \bz_{t-j} + \bvarepsilon_t,
\end{split}
\end{equation}
where $\btau^* = (\bI_N-\sum_{j=1}^p\bPhi_j)\btau$. The long-run relations are contained in the $N \times r$-matrix $\bB$, while the $N \times r$ matrix $\bA$ contains the corresponding loadings. Here the variable $r$ describes the number of cointegrating relations in the systems. If $r=0$, we adopt the convention that $\bA \bB^\prime = 0$; in this case $\bz_t$ is a pure $N$-dimensional unit root process. If $r=N$, all series are $I(0)$. To ensure that $\bz_t$ is at most an $I(1)$ process, the lag polynomial $\bC(z):= (1-z)-\bA \bB^\prime z - \sum_{j=1}^p \bPhi_j (1-z) z^j$ and matrices $\bA$ and $\bB$ should satisfy standard conditions that can be found in, inter alia, \citet{Johansen95}. Under these assumptions, exactly $N-r$ roots of the lag polynomial $\bC(z)$ are equal to unity, while the remaining $r$ roots lie outside the unit circle.

The typical interpretation of the VECM is that all series are $I(1)$, but $r$ linear combinations of the series are $I(0)$. However, it may also be the case that some individual series within the VECM are actually $I(0)$; these define ``trivial'' cointegration relations as any linear combination of these series remain $I(0)$. Thus the setup allows for observing a dataset with a mix of $I(0)$ and $I(1)$ series.

From the Granger Representation Theorem \citep[cf.][p.~49]{Johansen95}, we can obtain the \emph{common trend representation} of \eqref{eq:VECM}, which is given by
\begin{equation} \label{eq:CTR}
\bz_t = \bmu + \btau t + \bC \bs_t + \bu_t,
\end{equation}
where $\bC$ is an $N \times N$ matrix of rank $N-r$,\footnote{If $r=0$, we set $\bC = \bm 0$.} $\bs_t = \sum_{i=1}^t \bvarepsilon_t$ are the stochastic trends and $\bu_t$ is a stationary process. This representation show that $\bz_t$ can be decomposed in a deterministic process, an $I(1)$ part of common trends, $\bC \bs_t$, and a stationary part $\bu_t$.

To see the commonality of the trends, note that as $\bC$ is of reduced rank, we can define $N \times (N - r)$ matrices $\bLambda$ and $\bGamma$ such that $\bC = \bLambda \bGamma^\prime$. Then defining the $N - r \times 1$-vector $\bof_t = \bGamma^\prime \bs_t$, we can write \eqref{eq:CTR} as
\begin{equation} \label{eq:CFR}
\bz_t = \bmu + \btau t + \bLambda \bof_t + \bu_t.
\end{equation}
We can now see the common trends as \emph{common factors}, which provides a convenient way to think about cointegration in high dimensions.

This brings us to an alternative way to represent cointegration through a common factor structure from the outset. This form was considered by \citet{BaiNg04} among others to investigate different sources of nonstationarity in a panel data context. In this case we start from \eqref{eq:CFR}, assuming that the elements of both $\bof_t$ and $\bu_t$ can be $I(0)$ or $I(1)$. The combination of the two then determines the properties of the series $\bz_t$. Consider a single series $\bz_{i,t}$, which can be represented as
\begin{equation*}
\bz_{i,t} = \mu_i + \tau_i t + \blambda_i^\prime \bof_t + u_{i,t},
\end{equation*}
where $\blambda_i^\prime$ denotes the $i$-th row of $\bLambda$. Note that $\bz_{i,t}$ is $I(0)$ only if both $\bu_{i,t}$ and $\blambda_i^\prime \bof_t$ are $I(0)$, where the latter occurs if either all factors $\bof_t$ are $I(0)$, or no $I(1)$ factors load on series $i$. Similarly, cointegration between series $i$ and $j$ requires that both $u_{i,t}$ and $u_{j,t}$ are $I(0)$.

\begin{remark}
For expositional simplicity we do not consider $I(2)$ variables here. While the VECM can be extended to allow for $I(2)$ series, see e.g.~\citet{Johansen95ET}, in practice most cointegration analyses are performed on $I(1)$ series. If the data contains (suspected) $I(2)$ series, these are generally differenced before commencing the cointegration analysis.

Similarly, one could think of the data generating process (DGP) as being of infinite lag order, rather than fixed order $p$. In this case the VECM with fixed order can be thought of as an approximation to the infinite order model, where $p$ should be large enough to capture ``enough'' of the serial correlation. Either way, in applications $p$ is generally not known and has to be estimated.
\end{remark}

\section{Transformations to Stationarity and Unit Root Pre-Testing} \label{sec:trans}
In this section we discuss how to determine the appropriate transformations ---in particular how often the series need to be differenced--- in order to obtain only stationary time series in our dataset. While established datasets, such as the FRED-MD, come with an overview of the appropriate transformation for each series, this is generally not the case and data-driven methods are needed. Thus, one normally has to apply unit root or stationarity tests to determine the order of integration, and the corresponding transformation. In this section we investigate how to approach this pre-testing problem.

First, we investigate unit root tests in more detail, and highlight some of their characteristics that one should take into account when considering high-dimensional macroeconomic forecasting. Second, we discuss how to deal with the multiple testing problem that arises from the fact that we need to combine unit root tests on many time series.

\subsection{Unit Root Test Characteristics} \label{sec:urtest}
Even though the literature on unit root testing has grown exponentially since the seminal paper of \citet{DickeyFuller79}, discussing at length the characteristics of various unit root tests, unit root pre-testing is often done in an automatic, routine-like, way by considering classical tests such as augmented Dickey-Fuller (ADF) tests. However, these tests have various problematic characteristics which may accumulate when applied in high-dimensional problems. While we cannot discuss all of these here, let us briefly mention some of particular relevance for macroeconomic forecasting. An extensive overview of unit root testing is provided by \citet{Choi15}.\footnote{Given the greater popularity of tests where the null hypothesis is a unit root over tests with stationarity as the null, we focus exclusively on unit root tests here. However, most of the discussion applies to stationarity tests as well.}

\subsubsection{Size distortions} \label{sec:ursz}
Standard unit root tests are very prone to size distortions. One source is neglected serial correlation \citep[cf.][]{Schwert89}, while another is time-varying volatility \citep{Cavaliere05}. For both sources, bootstrap methods have proven a successful means to counteract the size distortions; however, while for serial correlation any ``off-the-shelf'' time series bootstrap method can be used \citep[see][for an overview and comparison]{PSU08}, dealing with general forms of heteroskedasticity requires a unit root test based on the wild bootstrap \citep{CavaliereTaylor08, CavaliereTaylor09ER}.

It should be noted that unconditional volatility changes pose a particular concern for macroeconomic time series. Many datasets such as FRED-MD span the period of the Great Moderation, which has significantly affected the volatility of macroeconomic time series \citep{JustinianoPrimiceri08,StockWatson03}. It would therefore appear wise to take potential volatility changes into account when selecting an appropriate unit root test.

\subsubsection{Power and specification considerations} \label{sec:urpw}
The power properties of the different unit root tests proposed vary considerably, and generally optimal tests do not exist. One particular source of variation is the magnitude of the initial condition, where for instance the DF-GLS test of \citet{ERS96} is optimal when the initial condition is zero, but the ADF test is much more powerful when the initial condition is large \citep{MuellerElliott03}. An even larger source of variation is the presence or absence of a deterministic trend. Unit root tests with a trend included (or, equivalently, unit root tests performed on detrended data) are considerably less powerful than without trend (performed on demeaned data). On the other hand, if a trend is not included when the data do contain one, the unit root test is not correctly sized anymore \citep{HLT09}.

While dealing with such issues is manageable in unit root testing for a single series, this changes when considering large datasets. For instance, deciding whether to include a trend in the unit root test can be based on a combination of theory, visual inspection, pre-testing, and comparing outcomes of different tests with or without a trend. However, such an analysis has to be done manually for each series involved, which quickly becomes problematic if the dimension of the dataset increases. This is even more problematic for modern Big Data sets, such as Google Trends, for which no theory exists to guide the practitioner, and where the dimension can become arbitrarily large.

As such one would like to have an automatic way of choosing good specifications for the unit root tests, that may differ across series. One easy way is provided by the union of unit root tests principle proposed by \citet{HLT09,HLT12}, in which several unit root tests are performed, and the unit root null hypothesis is rejected if one of the tests rejects (when corrected for multiple testing). In particular, \citet{HLT12} consider a union of the ADF and DF-GLS tests, both with and without linear trend, to cover uncertainty about both trend and initial condition. \citet{SmeekesTaylor12} consider a wild bootstrap version of this test that is robust to time-varying volatility. The test statistic for series $i$ takes the form
\begin{equation}\label{eq:URunion}
\begin{split}
UR_i =\min &\left(\left( \frac{x_i} {c_{i,GLS}^{\mu*} (\alpha)} \right) GLS_i^{\mu}, \left( \frac{x_i} {c_{i,GLS}^{\tau*} (\alpha)} \right) GLS_i^{\tau},\right.\\
&\quad \left. \left( \frac{x_i} {c_{i,ADF}^{\mu*} (\alpha)} \right) ADF_i^{\mu}, \left( \frac{x_i} {c_{i,ADF}^{\tau*} (\alpha)} \right) ADF_i^{\tau} \right),
\end{split}
\end{equation}
where $ADF_i$ and $GLS_i$ are the ADF and DF-GLS test performed on series $i$, while superscript $\mu$ and $\tau$ indicate whether the series are demeaned or detrended respectively. The bootstrap critical values such as $c_{i,GLS}^{\mu*} (\alpha)$ used in the scaling factors are determined in a preliminary bootstrap step as the individual level $\alpha$ critical values of the four tests. The variable $x_i$ is a scaling factor to which the statistics are scaled. Any $x_i < 0$ suffices to preserve the left-tail rejection region; if one additionally takes $x_i$ the same value for all series $i$, test statistics become comparable across series, which facilitates the multiple comparisons discussed in the next subsection.

\subsection{Multiple Unit Root Tests} \label{sec:multur}
Performing a unit root test for every series separately raises issues associated with multiple testing. In particular, the probability of incorrect classifications rises with the number of tests performed. If each test has a significance level of 5\%, we may also expect roughly 5\% of the $I(1)$ series to be incorrectly classified as $I(0)$. In a high-dimensional dataset this can quickly lead to a significant number of incorrectly classified series. It will of course depend on the specific application whether this is problematic ---a priori we cannot say whether the ``important'' series will be correctly classified or not--- but to avoid such issues one can formally account for multiple testing.

There is a huge statistical literature about multiple testing; \citet{RSW08} provide an overview with a focus on econometric applications. Here we briefly discuss the most prominent methods developed for the purposes of unit root testing. Before discussing the different methods to control for multiple testing, let us set up the general framework. Let $UR_1, \ldots, UR_N$ denote the unit root test statistics for series 1 up to $N$, assuming they reject for small values of the statistics.\footnote{We can assume this without loss of generality as any test statistic can be modified to indeed do so.} It is important to choose the test statistics such that they are directly comparable, in the sense that their marginal distributions are the same. If this is the case, then the ranking
\begin{equation} \label{eq:ranking}
UR_{(1)} \leq \ldots \leq UR_{(R)} \leq UR_{(R+1)} \leq \ldots \leq UR_{(N)},
\end{equation}
where $UR_{(i)}$ denotes the $i$-th order statistic of $UR_1, \ldots, UR_N$, corresponds to a ranking from ``most significant'' to ``least significant''. To ensure the comparability of the test statistics, one needs to eliminate nuisance parameters from their distribution. Hence, simply using the bootstrap to absorb nuisance parameters is not sufficient; instead, one often needs to transform (for instance to $p$-values) or scale the statistics appropriately. In the union tests of \eqref{eq:URunion}, the scaling is done automatically by setting $x_i=-1$ for all units.

Given the ranking in \eqref{eq:ranking}, the objective is to find an appropriate cut-off point $R$ such that for all statistics less than or equal $UR_{(R)}$ the unit root hypothesis is rejected, and for all statistics larger it is not rejected. How this threshold is determined depends on how multiple testing is controlled for.

\subsubsection{Controlling generalized error rates} \label{sec:ger}
\emph{Generalized error rates} provide multivariate extensions of the standard Type I error. The most common is the \emph{familywise error rate (FWE)}, which is defined as the probability of making at least one false rejection of the null hypothesis. This can easily be controlled by the popular Bonferroni correction. However, this is very conservative as it is valid under any form of dependence. On the contrary, if the bootstrap is used to capture the actual dependence structure among the tests, one can control for multiple testing without the need for being conservative. This approach is followed by \citet{Hanck09}, who controls FWE in unit root testing by applying the bootstrap algorithm proposed by \citet{RomanoWolf05}.

While controlling FWE makes sense when $N$ is small, in typical high-dimensional datasets FWE becomes too conservative. Instead, one can control the \emph{false discovery rate (FDR)} originally proposed by \citet{BenjaminiHochberg95}, which is defined as 
\begin{equation*}
FDR = \E \left[\frac{F}{R} \mathbbm{1} (R > 0)\right],
\end{equation*}
where $R$ denote the total number of rejections, and $F$ the number of false rejections. The advantage of the FDR is that it scales with increasing $N$, and thus is more appropriate for large datasets. However, most non-bootstrap methods are either not valid under arbitrary dependence or overly conservative. \citet{MoonPerron12} compare several methods to control FDR and find that the bootstrap method of \citet{RSW08fdr}, hereafter denoted as BFDR, does not share these disadvantages and clearly outperforms the other methods. A downside of this method however is that the algorithm is rather complicated and time-consuming to implement. Globally, the algorithm proceeds in a sequential way by starting to test the ``most significant'' series, that is, the smallest unit root test statistic. This statistic is then compared to an appropriate critical values obtained from the bootstrap algorithm, where the bootstrap evaluates all scenarios possible in terms of false and true rejections given the current progression of the algorithm. If the null hypothesis can be rejected for the current series, the algorithm proceeds to the next most significant statistic and the procedure is repeated. Once a non-rejection is observed, the algorithm stops. For details we refer to \citet{RSW08fdr}. This makes the bootstrap FDR method a \emph{step-down} method, contrary to the original \citet{BenjaminiHochberg95} approach which is a step-up method starting from the least significant statistic.

\subsubsection{Sequential Testing} \label{sec:seqtest}
\citet{Smeekes15} proposes an alternative bootstrap method for multiple unit root testing based on sequential testing. In a first step, the null hypothesis that all $N$ series are $I(1)$ --hence $p_1=0$ series are $I(0)$--- is tested against the alternative that (at least) $p_2$ series are $I(0)$. If the null hypothesis is rejected, the $p_2$ most significant statistics in \eqref{eq:ranking} are deemed $I(0)$ and removed from consideration. Then the null hypothesis that all remaining $N - p_2$ series are $I(1)$ is tested against the alternative that at least $p_2$ of them are $I(0)$, and so on. If no rejections are observed, the final rounds tests $p_K$ $I(0)$ series against the alternative of $N$ $I(0)$ series. The numbers $p_2, \ldots, p_K$ as well as the number of tests $K$ are chosen by the practitioner based on the specific application at hand. By choosing the numbers as $p_k = [q_k N]$, where $q_1, \ldots, q_K$ are desired quantiles, the method automatically scales with $N$.

Unlike the BFDR method, this Bootstrap Sequential Quantile Test (BSQT) is straightforward and fast to implement. However, it is dependent on the choice of numbers $p_k$ to be tested; its ``error allowance'' is therefore of different nature than error rates like FDR. \citet{Smeekes15} shows that, when $p_J$ units are found to be $I(0)$, the probability that the true number of $I(0)$ series lies outside the interval $[p_{J-1}, p_{J+1}]$ is at most the chosen significance level of the test. As such, there is some uncertainty around the cut-off point. 

It might therefore be tempting to choose $p_k = k-1$ for all $k=1, \ldots, N$, such that this uncertainty disappears. However, as discussed in \citet{Smeekes15}, applying the sequential method to each series individually hurts power if $N$ is large as it amounts to controlling FWE. Instead, a better approach is to iterate the BSQT method; that is, it can be applied in a second stage just to the interval $[p_{J-1}, p_{J+1}]$ to reduce the uncertainty. This can be iterated until few enough series remain to be tested individually in a sequential manner. On the other hand, if $p_1, \ldots, p_K$ are chosen sensibly and not spaced too far apart, the uncertainty is limited to a narrow range around the ``marginally significant'' unit root tests. These series are at risk of miss-classification anyway, and the practical consequences of incorrect classification for these series on the boundary of a unit root are likely small.

\citet{Smeekes15} performs a Monte Carlo comparison of the BSQT and BFDR methods, as well as several methods proposed in the panel data literature such as \citet{Ng08} and \citet{ChortareasKapetanios09}. Globally BSQT and BFDR clearly outperform the other methods, where BFDR is somewhat more accurate than BSQT when the time dimension $T$ is at least of equal magnitude as the number of series $N$. On the other hand, when $T$ is much smaller than $N$ BFDR suffers from a lack of power and BSQT is clearly preferable. In our empirical applications we will therefore consider both BFDR and BSQT, as well as the strategy of performing individual tests without controlling for multiple testing. 

\begin{remark}
An interesting non-bootstrap alternative is the panel method proposed by \citet{PVWW15}, which has excellent performance in finite samples. However, implementation of this method requires that $T$ is strictly larger than $N$, thus severely limiting its potential for analyzing Big Data. Another alternative would be to apply the model selection approach through the adaptive lasso by \citet{Kock16} which avoids testing all together. However, this has only been proposed in a univariate context and its properties are unknown for the type of application considered here.
\end{remark}

\subsubsection{Multivariate Bootstrap Methods} \label{sec:mboot}
All multiple testing methods described above require a bootstrap method that can not only account for dependence within a single time series, but can also capture the dependence structures between series. Accurately modelling the dependence between the individual test statistics is crucial for proper functioning of the multiple testing corrections. Capturing the strong and complex dynamic dependencies between macroeconomic series requires flexible bootstrap methods that can handle general forms of dependence.

\citet{MoonPerron12} and \citet{Smeekes15} use the moving-blocks bootstrap (MBB) based on the results of \citet{PSU11} who prove validity for mixed $I(1)/I(0)$ panel datasets under general forms of dependence. However the MBB has two disadvantages. First, it can only be applied to balanced datasets where each time series is observed over the same period. This makes application to datasets such as FRED-MD difficult, at least without deleting observations for series that have been observed for a longer period. Second, the MBB is sensitive to unconditional heteroskedasticity, which makes its application problematic for series affected by the Great Moderation.

\emph{Dependent wild bootstrap (DWB)} methods address both issues while still being able to capture complex dependence structure. Originally proposed by \citet{Shao10} for univariate time series, they were extended to unit root testing by \citet{SmeekesUrbain14RM} and \citet{RhoShao19}, where the former paper considers the multivariate setup needed here. A general wild bootstrap algorithm for multivariate unit root testing looks as follows:
\begin{enumerate}
\item Detrend the series $\{\bz_t\}$ by OLS; that is, let $\hat{\bzeta}_t = (\widehat{\zeta}_{1,t}, \ldots, \widehat{\zeta}_N )^\prime$ where
\begin{equation*}
\widehat{\zeta}_{i,t} = z_{i,t} - \widehat{\mu}_i - \widehat{\tau}_i t, \qquad i = 1, \ldots, N, \quad t = 1, \ldots, T
\end{equation*}
and $(\widehat{\mu}_i, \widehat{\tau}_i)^\prime$ are the OLS estimators of $(\mu_i, \tau_i)^\prime$.

\item Transform $\widehat{\bzeta}_t$ to a multivariate $I(0)$ series $\widehat{\bu}_t = (\widehat{u}_{1,t}, \ldots, \widehat{u}_{N,t})^\prime$ by setting
\begin{equation*}
\widehat{u}_{i,t} = \widehat{\zeta}_{i,t} - \widehat{\rho}_i \widehat{\zeta}_{i,t-1}, \qquad i = 1, \ldots, N, \quad t = 1, \ldots, T,
\end{equation*}
where $\widehat{\rho}_i$ is either an estimator of the largest autoregressive root of $\{\widehat{\zeta}_{i,t}\}$ using for instance an (A)DF regression, or $\widehat{\rho}_i = 1$.

\item Generate a univariate sequence of \emph{dependent} random variables $\xi_1^*, \ldots, \xi_N^*$ with the properties that $\E^* \xi_t^* = 0$ and $\E^* \xi_t^{*2} = 1$ for all $t$. Then construct bootstrap errors $\bu_t^* = (u_{1,t}^*, \ldots, u_{N,t}^*)^\prime$ as
\begin{equation} \label{eq:DWB}
u_{i,t}^* = \xi_t^* \hat{u}_{i,t}, \qquad i = 1, \ldots, N, \quad t = 1, \ldots, T.
\end{equation}

\item Let $\bz_t^* = \sum_{s=1}^t \bu_s^*$ and calculate the desired unit root test statistics $UR_1^*, \ldots, UR_N^*$ from $\{\bz_t^*\}$. Use these bootstrap test statistics in an appropriate algorithm for controlling multiple testing.
\end{enumerate}

Note that, unlike for the MBB, in \eqref{eq:DWB} no resampling takes place, and as such missing values ``stay in their place'' without creating new ``holes'' in the bootstrap samples. This makes the method applicable to unbalanced panels. Moreover, heteroskedasticity is automatically taken into account by virtue of the wild bootstrap principle. Serial dependence is captured through the dependence of $\{\xi_t^*\}$, while dependence across series is captured directly by using the same, univariate, $\xi_t^*$ for each series $i$. \citet{SmeekesUrbain14RM} provide theoretical results on the bootstrap validity under general forms of dependence and heteroskedasticity.

There are various options to draw the dependent $\{\xi_t^*\}$; \citet{Shao10} proposes to draw these from a multivariate normal distributions, where the covariance between $\xi_s^*$ and $\xi_t^*$ is determined by a kernel function with as input the scaled distance $\abs{s-t}/\ell$. The tuning parameter $\ell$ serves as a similar parameter as the block length in the MBB; the larger it is, the more serial dependence is captured. \citet{SmeekesUrbain14RM} and \citet{FSU18} propose generating $\{\xi_t^*\}$ through an AR(1) process with normally distributed innovations and AR parameter $\gamma$, where $\gamma$ is again a tuning parameter that determines how much serial dependence is captured. They label this approach the autoregressive wild bootstrap (AWB), and show that the AWB generally performs at least as well as \citepos{Shao10} DWB in simulations.

Finally, one might consider the sieve wild bootstrap used in \citet{CavaliereTaylor09ER} and \citet{SmeekesTaylor12}, where the series $\{\widehat{\bu}_t\}$ are first filtered through individual AR processes, and the wild bootstrap is applied afterwards to the residuals. However, as \citet{SmeekesUrbain14} show that this method cannot capture complex dynamic dependencies across series, it should not be used in this multivariate context. If common factors are believed to be the primary source of dependence across series, factor bootstrap methods such as those considered by \citet{Trapani13} or \citet{GoncalvesPerron14} could be used as well.

\section{High-Dimensional Cointegration} \label{sec:coint}

In this section, we discuss various recently proposed methods to model high-dimensional (co)integrated datasets. Similar to the high-dimensional modelling of stationary datasets, two main modelling approaches can be distinguished. One approach is to summarize the complete data into a much smaller and more manageable set through the extraction of common factors and their associated loadings, thereby casting the problem into the framework represented by \eqref{eq:CFR}. Another approach is to consider direct estimation of a system that is fully specified on the observable data as in \eqref{eq:VECM}, under the implicit assumption that the true DGP governing the long- and short-run dynamics is sparse, i.e. the number of non-zero coefficients in said relationships is small. These two approaches, however, rely on fundamentally different philosophies and estimation procedures, which constitute the topic of this section.\footnote{Some recent papers such as \citet{OnatskiWang18} and \citet{ZRY18} have taken different, novel approaches to high-dimensional cointegration analysis. However, these methods do not directly lend themselves to forecasting and are therefore not discussed.}

\subsection{Modelling Cointegration through Factor Structures}

Factor models are based on the intuitive notion that all variables in an economic system are driven by a small number of common shocks, which are often thought of as representing broad economic phenomena such as the unobserved business cycle. On (transformed) stationary macroeconomic data sets, the extracted factors have been successfully applied for the purpose of forecasting by incorporating them in dynamic factor models \citep{FHL05}, factor-augmented vector autoregressive (FAVAR) models \citep{BBE05} or  single-equation models \citep{StockWatson02JASA,StockWatson02JBES}. Recent proposals are brought forward in the literature that allow for application of these techniques on non-stationary and possibly cointegrated datasets. In Section \ref{sec:DFM} the dynamic factor model proposed by \citet{BLL17,BLL18} is discussed and Section \ref{sec:FECM} details the factor-augmented error correction model by \citet{BMM14,BMM16}. As both approaches require an a priori choice on the number of common factors, we briefly discuss estimation of the factor dimension in Section \ref{sec:f_dim}

\subsubsection{Dynamic Factor Model}\label{sec:DFM}

A popular starting point for econometric modelling involving common shocks is the specification of a dynamic factor model. Recall our representation of an individual time series by
\begin{equation}\label{eq:CFT_i}
    z_{i,t} = \mu_i + \tau_it + \bm{\lambda}_i^\prime\bm{f}_t + u_{i,t},
\end{equation}
where $\bm{f}_t$ is the $N-r \times 1$ dimensional vector of common factors. Given a set of estimates for the unobserved factors, say $\hat{\bm{f}}_t$ for $t=1,\ldots,T$, one may directly obtain estimates for the remaining parameters in \eqref{eq:CFT_i} by solving the least-squares regression problem\footnote{Typically, the estimation procedure for $\hat{\bm{f}}_t$ provides the estimates $\hat{\bm{\Lambda}}$ as well, such that only the coefficients regulating the deterministic specification ought to be estimated.}
\begin{equation}
    \left(\hat{\bm{\mu}},\hat{\bm{\tau}},\hat{\bm{\Lambda}}\right) = \underset{\bm{\mu},\bm{\tau},\bm{\Lambda}}{\text{arg min}} \sum_{t=1}^T \left(\bm{z}_t - \bm{\mu} - \bm{\tau}\bm{t} - \bm{\Lambda}\hat{\bm{f}}_t\right)^2.
\end{equation}
The forecast for the realization of an observable time series at time period $T+h$ can then be constructed as
\begin{equation}\label{eq:Zit_forecast}
    \hat{z}_{i,T+h|T} = \hat{\mu}_i + \hat{\tau}_i (T+h) + \hat{\bm{\lambda}}_i\hat{\bm{f}}_{T+h|T}.
\end{equation}
This, however, requires the additional estimate $\hat{\bm{f}}_{T+h|T}$, which may be obtained through an explicit dynamic specification of the factors.

\citet{BLL18} assume that the differenced factors admit a reduced-rank vector autoregressive (VAR) representation, given by
\begin{equation}\label{eq:DFM}
    \bm{S}(L)\Delta \bm{f}_t = \bm{C}(L)\bm{\nu}_t,
\end{equation}
where $\bm{S}(L)$ is an invertible $N-r \times N-r$ matrix polynomial and $\bm{C}(L)$ is a finite degree $N-r \times q$ matrix polynomial. Furthermore, $\bm{\nu}_t$ is a $q \times 1$ vector of white noise common shocks with $N-r > q$. Inverting the left-hand side matrix polynomial and summing both sides, gives rise to the specification
\begin{equation}\label{eq:FT}
    \bm{f}_t = \bm{S}^{-1}(L)\bm{C}(L)\sum_{t=1}^T\bm{\nu}_t = \bm{U}(L)\sum_{t=1}^T\bm{\nu}_t = \bm{U}(1)\sum_{t=1}^T\bm{\nu}_t + \bm{U}^*(L)\bm{\nu}_t,
\end{equation}
where the last equation follows from application of the Beveridge-Nelson decomposition to $\bm{U}(L) = \bm{U}(1) + \bm{U}^*(L)(1-L)$. Thus, \eqref{eq:FT} reveals that the factors are driven by a set of common trends and stationary linear processes. Crucially, the assumption that the number of common shocks is strictly smaller than the number of integrated factors, i.e. $\bm{f}_t$ is a singular stochastic vector, implies that $\text{rank}\left(\bm{U}(1)\right) = q-d$ for $0\leq d < q$. Consequently, there exists a full column rank matrix $\bm{B}_f$ of dimension $N-r \times N-r-q+d$ with the property that $\bm{B}_f^\prime \bm{f}_t$ is stationary. Then, under the general assumption that the entries of $\bm{U}(L)$ are rational functions of $L$, \citet{BLL17} show that $\bm{f}_t$ admits a VECM representation of the form
\begin{equation}\label{eq:FT_VECM_inf}
    \Delta \bm{f}_t = \bm{A}_f\bm{B}_f^\prime \bm{f}_{t-1} + \sum_{j=1}^p \bm{G}_j \Delta\bm{f}_{t-j} + \bm{K}\bm{\nu}_t,
\end{equation}
where $\bm{K}$ is a constant matrix of dimension $N-r \times q$.

Since the factors in \eqref{eq:FT_VECM_inf} are unobserved, estimation of the system requires the use of a consistent estimate of the space spanned by $\bm{f}_t$. Allowing idiosyncratic components $\nu_{i,t}$ in \eqref{eq:CFT_i} to be either $I(1)$ or $I(0)$, and allowing for the presence of a non-zero constant $\mu_i$ and linear trend $\tau_i$, \citet{BLL18} propose an intuitive procedure that enables estimation of the factor space by the method of principal components. First, the data is de-trended with the use of a regression estimate:
\begin{equation*}
    \tilde{z}_{i,t} = z_{i,t} - \hat{\tau}_i t,
\end{equation*}
where $\hat{\tau}_i$ is the OLS estimator of the trend in the regression of $z_{i,t}$ on an intercept and linear trend. Then, similar to the procedure originally proposed by \citet{BaiNg04}, the factor loadings are estimated as $\hat{\bm{\Lambda}} = \sqrt{N}\hat{\bm{W}}$, where $\hat{\bm{W}}$ is the $N \times (N-r)$ matrix with normalized right eigenvectors of $T^{-1}\sum_{t=1}^T \Delta \tilde{\bm{z}}_t \Delta\tilde{\bm{z}}_t^\prime$ corresponding to the $N-r$ largest eigenvalues. The estimates for the factors are given by $\hat{\bm{f}}_t = \frac{1}{N}\hat{\bm{\Lambda}}^\prime \tilde{\bm{z}}_t$.

Plugging $\hat{\bm{f}}_t$ into \eqref{eq:FT_VECM_inf} results in
\begin{equation}\label{eq:FT_VECM}
    \Delta \hat{\bm{f}}_t = \bm{A}_f\bm{B}_f^\prime \hat{\bm{f}}_{t-1} + \sum_{j=1}^p \bm{G}_j \Delta\hat{\bm{f}}_{t-j} + \hat{\bm{\nu}}_t,
\end{equation}
which can be estimated using standard approaches, such as the maximum likelihood procedure proposed by \citet{Johansen95}. Afterwards, the iterated one-step-ahead forecasts $\Delta \hat{\bm{f}}_{T+1|T},\ldots,\Delta \hat{\bm{f}}_{T+h|T}$ are calculated from the estimated system, based on which the desired forecast $\hat{\bm{f}}_{T+h|T} = \hat{\bm{f}}_T + \sum_{k=1}^h \Delta \hat{\bm{f}}_{T+k|T}$ is obtained. The final forecast for $\hat{z}_{i,T+h|T}$ is then easily derived from \eqref{eq:Zit_forecast}.

\begin{remark}
Since the idiosyncratic components are allowed to be serially dependent or even $I(1)$, a possible extension is to explicitly model these dynamics. As a simple example, each $u_{i,t}$ could be modelled with a simple autoregressive model, from which the prediction $\hat{u}_{i,T+h|T}$ can be obtained following standard procedures \citep[e.g.][Ch. 4]{Hamilton94}. This prediction is then added to \eqref{eq:Zit_forecast}, leading to the final forecast
\begin{equation*}
    \hat{z}_{i,T+h|T} = \hat{\mu}_i + \hat{\tau}_i (T+h) + \hat{\bm{\lambda}}_i\hat{\bm{f}}_{T+h|T} + \hat{u}_{i,T+h|T}.
\end{equation*}
This extension leads to substantial improvements in forecast performance in the macroeconomic forecast application presented in Section \ref{sec:appl}.
\end{remark}

\subsubsection{Factor-Augmented Error Correction Model}\label{sec:FECM}

It frequently occurs that the variables of direct interest constitute only a small subset of the collection of observed variables. In this scenario, \citet{BMM14,BMM16,BMM17}, henceforth referred to as BMM, propose to model only the series of interest in a VECM system, while including factors extracted from the full dataset to proxy for the missing information from the excluded observed time series. 

The approach of BMM can be motivated starting from the common trend representation in \eqref{eq:CTR}. Partition the observed time series $\bm{z}_t = (\bm{z}_{A,t}^\prime,\bm{z}_{B,t}^\prime)^\prime$, where $\bm{z}_{A,t}$ is an $N_A \times 1$ vector containing the variables of interest. Then, we may rewrite \eqref{eq:CTR} as
\begin{equation}\label{eq:DFM_part}
    \begin{bmatrix}
    \bm{z}_{A,t}\\
    \bm{z}_{B,t}
    \end{bmatrix} = \begin{bmatrix}
    \bm{\mu}_A\\
    \bm{\mu}_B
    \end{bmatrix} + \begin{bmatrix}
    \bm{\tau}_A\\
    \bm{\tau}_B
    \end{bmatrix}t + \begin{bmatrix}
    \bm{\Lambda}_A\\
    \bm{\Lambda}_B
    \end{bmatrix}\bm{f}_t + \begin{bmatrix}
    \bm{u}_{A,t}\\
    \bm{u}_{B,t}
    \end{bmatrix}
\end{equation}
The idiosyncratic components in \eqref{eq:DFM_part} are assumed to be $I(0)$.\footnote{In principle, the proposed estimation procedure remains feasible in the presence of $I(1)$ idiosyncratic components. The theoretical motivation, however, relies on the concept of cointegration between the observable time series and a set of common factors. This only occurs when the idiosyncratic components are stationary.} Furthermore, both non-stationary $I(1)$ factors and stationary factors are admitted in the above representation. Contrary to \citet{BLL17}, BMM do not require the factors in \eqref{eq:DFM_part} to be singular.

To derive a dynamic representation better suited to forecasting the variables of interest, \citet{BMM14,BMM17} use the fact that when the subset of variables is of a lower dimension than the factors, i.e. $N_A > N-r$, $\bm{z}_{A,t}$ and $\bm{f}_t$ cointegrate. As a result, the Granger Representation Theorem implies the existence of an error correction representation of the form
\begin{equation}\label{eq:FECM_MA}
    \begin{bmatrix}
    \Delta \bm{z}_{A,t}\\
    \bm{f}_t
    \end{bmatrix} = \begin{bmatrix}
    \bmu_A\\
    \bmu_f
    \end{bmatrix} + \begin{bmatrix}
    \btau_A\\
    \btau_f
    \end{bmatrix}t + 
    \begin{bmatrix}
    \bm{A}_A\\
    \bm{A}_B
    \end{bmatrix}\bm{B}^\prime \begin{bmatrix}
    \bm{z}_{A,t-1}\\
    \bm{f}_{t-1}
    \end{bmatrix} + \begin{bmatrix}
    \bm{e}_{A,t}\\
    \bm{e}_{f,t}
    \end{bmatrix}.
\end{equation}
To account for serial dependence in \eqref{eq:FECM_MA}, \citet{BMM14} propose the approximating model
\begin{equation}\label{eq:FECM}
    \begin{bmatrix}
    \Delta \bm{z}_{A,t}\\
    \bm{f}_t
    \end{bmatrix} = \begin{bmatrix}
    \bmu_A\\
    \bmu_f
    \end{bmatrix} + \begin{bmatrix}
    \btau_A\\
    \btau_f
    \end{bmatrix}t + \begin{bmatrix}
    \bm{A}_A\\
    \bm{A}_B
    \end{bmatrix}\bm{B}^\prime \begin{bmatrix}
    \bm{z}_{A,t-1}\\
    \bm{f}_{t-1}
    \end{bmatrix} + \sum_{j=1}^p\bm{\Phi}_j\begin{bmatrix}
    \Delta \bm{z}_{A,t-j}\\
    \Delta \bm{f}_{t-j}
    \end{bmatrix} + \begin{bmatrix}
    \bm{\epsilon}_{A,t}\\
    \bm{\epsilon}_{f,t}
    \end{bmatrix},
\end{equation}
where the errors $\left(\bm{\epsilon}_{A,t}^\prime,\bm{\epsilon}_{f,t}^\prime\right)^\prime$ are $i.i.d.$

Similar to the case of the dynamic factor model in Section \ref{sec:DFM}, the factors in the approximating model \eqref{eq:FECM} are unobserved and need to be replaced with their corresponding estimates $\hat{\bm{f}}_t$. Under a set of mild assumptions, \citet{Bai04} shows that the space spanned by $\bm{f}_t$ can be consistently estimated using the method of principal components applied to the levels of the data. Assume that $\bm{f}_t = \left(\bm{f}_{ns,t}^\prime,\bm{f}_{s,t}^\prime\right)^\prime$ where $\bm{f}_{ns,t}$ and $\bm{f}_{s,t}$ contain $r_{ns}$ non-stationary and $r_s$ stationary factors, respectively. Let $\bm{Z}=(\bm{z}_1,\ldots,\bm{z}_T)$ be the $(N \times T)$ matrix of observed time series. Then, \citet{Bai04} shows that $\bm{f}_{ns,t}$ is consistently estimated by $\hat{\bm{f}}_{ns,t}$, representing the eigenvectors corresponding to the $r_{ns}$ largest eigenvalues of $\bm{Z}^\prime\bm{Z}$, normalized such that $\frac{1}{T^2}\sum_{t=1}^T \hat{\bm{f}}_{ns,t}\hat{\bm{f}}_{ns,t}^\prime = \bm{I}$. Similarly, $\bm{f}_{s,t}$ is consistently estimated by $\hat{\bm{f}}_{s,t}$, representing the eigenvectors corresponding to the next $r_s$ largest eigenvalues of $\bm{Z}^\prime\bm{Z}$, normalized such that $\frac{1}{T}\sum_{t=1}^T \hat{\bm{f}}_{s,t}\hat{\bm{f}}_{s,t}^\prime = \bm{I}$.

The final step in the forecast exercise consists of plugging in $\hat{\bm{f}}_t = \left(\hat{\bm{f}}_{ns,t}^\prime,\hat{\bm{f}}_{s,t}^\prime\right)^\prime$ into \eqref{eq:FECM}, leading to
\begin{equation}\label{eq:FECM_feasible}
    \begin{bmatrix}
    \Delta \bm{z}_{A,t}\\
    \hat{\bm{f}}_t
    \end{bmatrix} = \begin{bmatrix}
    \bmu_A\\
    \bmu_f
    \end{bmatrix} + \begin{bmatrix}
    \btau_A\\
    \btau_f
    \end{bmatrix}t + \begin{bmatrix}
    \bm{A}_A\\
    \bm{A}_B
    \end{bmatrix}\bm{B}^\prime \begin{bmatrix}
    \bm{z}_{A,t-1}\\
    \hat{\bm{f}}_{t-1}
    \end{bmatrix} + \sum_{j=1}^p\bm{\Phi}_j\begin{bmatrix}
    \Delta \bm{z}_{A,t-j}\\
    \Delta \hat{\bm{f}}_{t-j}
    \end{bmatrix} + \begin{bmatrix}
    \bm{\epsilon}_{A,t}\\
    \bm{\epsilon}_{f,t}
    \end{bmatrix}.
\end{equation}
Since in typical macroeconomic applications the number of factors is relatively small, feasible estimates for \eqref{eq:FECM_feasible} can be obtained from the maximum likelihood procedure of \citet{Johansen95}. The iterated one-step-ahead forecasts $\Delta \hat{\bm{z}}_{A,T+1|T},\ldots,\Delta \hat{\bm{z}}_{A,T+h|T}$ are calculated from the estimated system, which are then integrated to obtain the desired forecast $\hat{\bm{z}}_{A,T+h|T}$.

\subsubsection{Estimating the number of factors}\label{sec:f_dim}

Implementation of the factor models discussed in this section requires an a priori choice regarding the number of factors. A wide variety of methods to estimate the dimension of the factors is available. The dynamic factor model of \citet{BLL17,BLL18} adopts the estimation strategy proposed by \citet{BaiNg04}, which relies on first-differencing the data. Since, under the assumed absence of $I(2)$ variables, all variables in this transformed data set are stationary, the standard tools to determine the number of factors in the stationary setting are applicable. A non-exhaustive list is given by \citet{BaiNg02}, \citet{Hallin2007}, \citet{Alessi2010}, \citet{Onatski2010} and \citet{Ahn2013}.

The factor-augmented error correction model of \citet{BMM14,BMM16} adopts the estimation strategy proposed by \citet{BaiNg04}, which extracts the factors from the data in levels. While the number of factors may still be determined based on the differenced dataset, \citet{Bai04} proposes a set of information criteria that allows from estimation of the number of non-stationary factors without differencing the data. 

Conveniently, it is possible to combine factor selection procedures to separately determine the number of non-stationary and stationary factors. For example, the total number of factors, say $r_{ns}+r_s$, can be found based on the differenced dataset and one of the information criteria in \citet{BaiNg02}. Afterwards, the number of non-stationary factors, $r_{ns}$, is determined based on the data in levels using one of the the criteria from \citet{Bai04}. The number of stationary factors follows from the difference between the two criteria. Recently, \citet{BarigozziTrapani18} propose a novel approach to discern the number of $I(0)$ factors, zero-mean $I(1)$ factors, and factors with a linear trend. Their method however requires that all idiosyncratic components are $I(0)$.

\subsection{Sparse Models}

Rather than extracting common factors, an alternative approach to forecasting with macroeconomic data is full-system estimation with the use of shrinkage estimators \citep[e.g.][]{DeMol2008,Stock2012,Callot2014}. The general premise of shrinkage estimators is the so-called bias-variance trade-off, i.e. the idea that, by allowing a relatively small amount of bias in the estimation procedure, a larger reduction in variance may be attained. A number of shrinkage estimators, among which the lasso originally proposed by \citet{Tibshirani96}, simultaneously perform variable selection and model estimation. Such methods are natural considerations when it is believed that the data generating process is sparse, i.e. only a small subset of variables among the candidate set is responsible for the variation in the variables of interest. Obviously, such a viewpoint is in sharp contrast with the philosophy underlying the common factor framework. However, even in cases where a sparse data generating process is deemed unrealistic, shrinkage estimators can remain attractive due to their aforementioned bias-variance trade-off \citep{SmeekesWijler18}.

For expositional convencience, we assume in this section that either $\bm{\mu}$ and $\bm{\tau}$ are zero or that $\bm{z}_t$ is de-meaned and de-trended. Defining $\bm{\Pi} = \bm{A}\bm{B}^\prime$, model \eqref{eq:VECM} is then given by
\begin{equation*}
    \Delta \bm{z}_t = \bm{\Pi}\bm{z}_{t-1} + \sum_{j=1}^p \bm{\Phi}_j\Delta \bm{z}_{t-j} + \bm{\epsilon}_t,
\end{equation*}
which in matrix notation reads as
\begin{equation}\label{eq:VECM_matrix}
    \Delta \bm{Z} = \bm{\Pi}\bm{Z}_{-1} + \bm{\Phi}\Delta \bm{X} + \bm{E},
\end{equation}
where $\Delta \bm{Z} = \left(\Delta \bm{z}_1,\ldots,\Delta \bm{z}_T\right)$, $\bm{Z}_{-1} = \left(\bm{z}_0,\ldots,\bm{z}_{T-1}\right)$, $\bm{\Phi} = \left(\bm{\Phi}_1,\ldots,\bm{\Phi}_p\right)$ and  $\Delta \bm{X} = \left(\Delta \bm{x}_0,\ldots,\Delta \bm{x}_{T-1}\right)$, with $\bm{x}_t = \left(\bm{z}_t^\prime,\ldots,\bm{z}_{t-p+1}^\prime\right)^\prime$.

\subsubsection{Full-system estimation}

Several proposals to estimate \eqref{eq:VECM_matrix} with the use of shrinkage estimators are brought forward in recent literature. \cite{LiaoPhillips15} proposes an automated approach that simultaneously enables sparse estimation of the coefficient matrices $\left(\bm{\Pi},\bm{\Phi}\right)$, including the cointegrating rank of $\bm{\Pi}$ and the short-run dynamic lag order in $\bm{\Phi}$. However, while the method has attractive theoretical properties, the estimation procedure involves non-standard optimization over the complex plane and is difficult to implement even in low dimensions \citep[][p. 424]{LiangSchienle19}. Accordingly, we do not further elaborate on their proposed method, but refer the interested reader to the original paper.

\citet{LiangSchienle19} develop an automated estimation procedure that makes use of a QR-decomposition of the long-run coefficient matrix. They propose to first regress out the short-run dynamics, by post-multiplying \eqref{eq:VECM_matrix} with $\bm{M} = \bm{I}_T-\Delta\bm{X}^\prime\left(\Delta\bm{X}^\prime\Delta\bm{X}\right)^{-1}\Delta\bm{X}$, resulting in
\begin{equation}\label{eq:VECM_LRD}
    \Delta \tilde{\bm{Z}} = \bm{\Pi}\tilde{\bm{Z}}_{-1} + \tilde{\bm{E}},
\end{equation}
with $\Delta \tilde{\bm{Z}} = \Delta \bm{Z}M$, $\tilde{\bm{Z}}_{-1} = \bm{Z}_{-1}M$ and $\tilde{\bm{E}} = \bm{E}M$. The key idea behind the method proposed by \citeauthor{LiangSchienle19} is to decompose the long-run coefficient matrix into
\begin{equation*}
    \bm{\Pi} = \bm{Q}\bm{R},
\end{equation*}
where $\bm{Q}^\prime\bm{Q} = \bm{I}_N$ and $\bm{R}$ is an upper-triangular matrix. Such a representation can be be calculated from the $QR$-decomposition of $\bm{\Pi}$ with column pivoting. 

The column pivoting orders the columns in $\bm{R}$ according to size, such that zero elements occur at the ends of the rows. As a result, the rank of $\bm{\Pi}$ corresponds to the number of non-zero columns in $\bm{R}$. Exploiting this rank property requires an initial estimator for the long-run coefficient matrix, such as the OLS estimator
\begin{equation*}
    \Hat{\bm{\Pi}}_{OLS} = \left(\Delta \tilde{\bm{Z}} \Tilde{\bm{Z}}_{-1}^\prime \right)\left(\Tilde{\bm{Z}}_{-1}\Tilde{\bm{Z}}_{-1}^\prime\right)^{-1},
\end{equation*}
proposed by \citet{LiangSchienle19}. The $QR$-decomposition with column-pivoting is then calculated from $\Hat{\bm{\Pi}}_{OLS}^\prime$, resulting in the representation $\Hat{\bm{\Pi}}_{OLS} = \Hat{\bm{R}}_{OLS}^\prime\Hat{\bm{Q}}_{OLS}^\prime$.\footnote{As part of their theoretical contributions, \citet{LiangSchienle19} show that the first $r$ columns of $\Hat{\bm{Q}}$ consistently estimate the space spanned by the cointegrating vectors $\bm{B}$ in \eqref{eq:VECM}, in an asymptotic framework where the dimension $N$ is allowed to grow at rate $T^{1/4-\nu}$ for $\nu>0$.} Since the unrestricted estimator $\Hat{\bm{\Pi}}_{OLS}$ will be full-rank, $\Hat{\bm{R}}_{OLS}$ is a full-rank matrix as well. However, by the consistency of $\Hat{\bm{\Pi}}_{OLS}$ and the ordering induced by the column-pivoting step, the last $N-r$ columns are expected to contain elements that are small in magnitude. Accordingly, a well-chosen shrinkage estimator that penalizes the columns of $\bm{R}$ may be able to separate the relevant from the irrelevant columns.

Let $\Hat{\bm{R}} = \left(\hat{\bm{r}}_1,\ldots,\hat{\bm{r}}_N\right)$, $\Hat{r}_j = \left(\Hat{r}_{1,j},\ldots,\Hat{r}_{N,j}\right)^\prime$, $\norm{\Hat{\bm{r}}_j}_2 = \sqrt{\sum_{i=1}^N \Hat{r}_{i,j}^2}$ and $\Hat{\mu}_k = \sqrt{\sum_{i=k}^N \Hat{r}_{k,i}^2}$.
\begin{equation}\label{eq:SVECM_LS}
    \Hat{\bm{R}} = \underset{\bm{R}}{\text{arg min}} \norm{\Delta \bm{Z} - \bm{R}^\prime \Hat{\bm{Q}}^\prime \bm{Z}_{-1}}_2^2 + \lambda\sum_{j=1}^N\frac{\norm{\Hat{\bm{r}}_j}_2}{\Hat{\mu}_j},
\end{equation}
where $\lambda$ is a tuning parameter that controls the degree of regularization, with larger values resulting in more shrinkage. Weighting the penalty for each group by $\hat{\mu}_j$ puts a relatively higher penalty on groups for which the initial OLS estimates are small. The estimator clearly penalizes a set of pre-defined groups of coefficients, i.e. the columns of $\bm{R}$, and, therefore, is a variant of the group lasso for which numerous algorithms are available \citep[e.g.][]{Meier2008,Friedman2010,Simon2013}. The final estimate for the long-run coefficient matrix is obtained as $\Hat{\bm{\Pi}}=\Hat{\bR}^\prime\Hat{\bQ}_{OLS}^\prime$. 

The procedure detailed thus far focuses solely on estimation of the long-run relationships and requires an a priori choice of the lag order $p$. Furthermore, a necessary assumption is that initial OLS estimates are available, thereby restricting the admissible dimension of the system to $N(p+1) < T$. Within this restricted dimension, the short-run coefficient matrix $\bm{\Phi}$ can be consistently estimated by OLS and the corresponding lag order may be determined by standard information criteria such as the BIC. Alternatively, a second adaptive group lasso can be employed to obtain the regularized estimates $\Hat{\bm{\Phi}} = \left(\Hat{\bm{\Phi}}_1,\ldots,\Hat{\bm{\Phi}}_p\right)$, see \citet[][p. 425]{LiangSchienle19} for details. The lag order is then determined by the number of non-zero matrices $\Hat{\bm{\Phi}}_i$ for $i \in \{1,\ldots,p\}$.

\citet{WilmsCroux16} propose a penalized maximum likelihood estimator to estimate sparse VECMs. Instead of estimating the cointegrating rank and coefficient matrices for a fixed lag order, the method of \citeauthor{WilmsCroux16} enables joint estimation of the lag order and coefficient matrices for a given cointegrating rank. Additionally, the penalized maximum likelihood procedure does not require the availability of initial OLS estimates and, therefore, notwithstanding computational constraints, can be applied to datasets of arbitrary dimension. Under the assumption of multivariate normality of the errors, i.e. $\bm{\epsilon}_t \sim \mathbb{N}\left(\bm{0},\bm{\Sigma}\right)$, the penalized negative log-likelihood is given by
\begin{equation}\label{eq:NLL}
\begin{split}
    \mathcal{L}\left(\bA,\bB,\bPhi,\bOmega\right) &= \frac{1}{T}\text{tr}\left((\Delta \bZ - \bA\bB^\prime\bZ_{-1} - \bPhi\Delta\bX)^\prime\bOmega(\Delta \bZ - \bA\bB^\prime\bZ_{-1} - \bPhi\Delta\bX)\right)\\
    &\quad -\log\abs{\bOmega} + \lambda_1P_1(\bB) + \lambda_2P_2(\bPhi) + \lambda_3P_3(\bOmega),
\end{split}
\end{equation}
where $\bOmega = \bSigma^{-1}$, and $P_1$, $P_2$ and $P_3$ being three penalty functions. The cointegrating vectors, short-run dynamics, and covariance matrix are penalized as
\begin{equation*}
    P_1(\bB) = \sum_{i=1}^N\sum_{j=1}^r \abs{\beta}_{i,j}, \quad P_2(\bPhi) = \sum_{i=1}^N\sum_{j=1}^{Np} \abs{\phi_{i,j}}, \quad P_3(\bOmega) = \sum_{i,j=1, i\neq j}^N \abs{\omega_{i,j}},
\end{equation*}
respectively. The use of $L_1$-penalization enables some elements to be estimated as exactly zero. The solution that minimizes \eqref{eq:NLL} is obtained through an iterative updating scheme, where the solution for a coefficient matrix is obtained by minimizing the objective function conditional on the remaining coefficient matrices. The full algorithm is described in detail in \citet[][p. 1527-1528]{WilmsCroux16} and R code is provided by the authors online.\footnote{\url{https://feb.kuleuven.be/public/u0070413/SparseCointegration/}}

\subsubsection{Single-equation estimation}\label{sec:SE}

Frequently, the forecast exercise is aimed at forecasting a small number of time series based on a large number of potentially relevant variables. The means of data reduction thus far considered utilize either data aggregation or subset selection. However, in cases where the set of target variables is small, a substantial reduction in dimension can be obtained through the choice of appropriate single-equation representations for each variable separately. 

\citet{SmeekesWijler18specs} propose the Penalized Error Correction Selector (SPECS) as an automated single-equation modelling procedure on high-dimensional (co)integrated datasets. Assume that the $N$-dimensional observed time series admits the decomposition $\bm{z}_t = (y_t,\bm{x}_t^\prime)^\prime$, where $y_t$ is the variable of interest and $\bm{x}_t$ are variables that are considered as potentially relevant in explaining the variation in $y_t$. Starting from the VECM system \eqref{eq:VECM_matrix}, a single-equation representation for $\Delta y_t$ can be obtained by conditioning on the contemporaneous differences $\Delta \bm{x}_t$. This results in
\begin{equation}\label{eq:SEM}
    \Delta y_t = \bdelta^\prime \bm{z}_{t-1} + \bpi^\prime \bw_t + \epsilon_{y,t},
\end{equation}
where $\bw_t = (\Delta \bx_t^\prime,\Delta \bz_{t-1}^\prime,\ldots,\Delta\bz_{t-p}^\prime)^\prime$\footnote{Details regarding the relationship between the components of the single-equation model \eqref{eq:SEM} and the full system \eqref{eq:VECM} are provided in \citet[][p. 5]{SmeekesWijler18specs}.}. The number of parameters to be estimated in the single-equation model \eqref{eq:SEM} is reduced to $N(p+2)-1$ as opposed to the original $N^2(p+1)$ parameters in \eqref{eq:VECM_matrix}. Nonetheless, for large $N$ the total number of parameters may still be too large to estimate precisely by ordinary least squares, if possible at all. Therefore, \citeauthor{SmeekesWijler18specs} propose a shrinkage procedure defined as
\begin{equation}\label{eq:SPECS}
\begin{split}
    \Hat{\bdelta},\Hat{\bpi} &= \underset{\bdelta,\bpi}{\text{arg min}} \sum_{t=1}^T \left(\Delta y_t - \bdelta^\prime \bz_{t-1} + \bpi^\prime \bw_t\right)^2 + P_\lambda(\bdelta,\bpi).
\end{split}
\end{equation}
The penalty function takes on the form
\begin{equation}
    P_\lambda(\bdelta,\bpi) = \lambda_G\norm{\bdelta} + \lambda_\delta \sum_{i=1}^N\omega_{\delta,i}^{k_\delta}\abs{\delta_i} + \lambda_\pi \sum_{j=1}^{N(p+1)-1}\omega_{\pi,j}^{k_\pi}\abs{\pi_j},
\end{equation}
where $\omega_{\delta,i}^{k_\delta} = 1/\abs{\hat{\delta}_{Init,i}}^{k_\delta}$ and $\omega_{\pi,j}^{k_\pi} = 1/\abs{\hat{\pi}_{Init,j}}^{k_\pi}$, with $\Hat{\bdelta}_{Init}$ and $\Hat{\bpi}_{Init}$ being some consistent initial estimates, such as OLS or ridge estimates. The tuning parameters $k_\delta$ and $k_\pi$ regulate the degree to which the initial estimates affect the penalty weights.

SPECS simultaneously employs individual penalties on all coefficients and a group penalty on $\bdelta$, the implied cointegrating vector. Absent of cointegration, this cointegrating vector is equal to zero, in which case the group penalty promotes the removal of the lagged levels as a group.\footnote{From a theoretical point of view, the group penalty is not required for consistent selection and estimation of the non-zero coefficients.} In the presence of cointegration, however, the implied cointegrating vector may still contain many zero elements. The addition of the individual penalties allow for correct recovery of this sparsity pattern. This combination of penalties is commonly referred to as the sparse group lasso and R code is provided by the authors.\footnote{\url{https://sites.google.com/view/etiennewijler/code?authuser=0}}

In the single-equation model, the variation in $y_t$ is explained by contemporaneous realizations of the conditioning variables $\bx_t$. Therefore, forecasting the variable of interest requires forecasts for the latter as well, unless their realizations become available to the researcher prior to the realizations of $y_t$. SPECS is therefore highly suited to nowcasting applications. While not originally developed for the purpose of forecasting, direct forecasts with SPECS can be obtained by modifying the objective function to
\begin{equation*}
\begin{split}
   \sum_{t=1}^T \left(\Delta_h y_t - \bdelta^\prime \bz_{t-1} + \bpi^\prime \bw_t\right)^2 + P_\lambda(\bdelta,\bpi),
\end{split}
\end{equation*}
where $\Delta_h y_t = y_{t+h}-y_t$. The direct $h$-step ahead forecast is then simply obtained as
$\Hat{y}_{T+h|T} = y_T + \Hat{\bdelta}^\prime \bz_{T-1} + \Hat{\bpi}^\prime \bw_T$.

\section{Empirical Applications} \label{sec:appl}
In this section we evaluate the methods discussed in Sections \ref{sec:trans} and \ref{sec:coint} in two empirical applications. First we forecast several US macroeconomic variables using the FRED-MD dataset of \citet{McCrackenNg16}. The FRED-MD dataset is a well-established and popular source for macroeconomic forecasting, and allows us to evaluate the methods in an almost controlled environment. Second we consider nowcasting Dutch unemployment using Google Trends data on frequencies of unemployment-related queries. This application not only highlights the potential of novel Big Datasets for macroeconomic purposes, but also puts the methods to the test in a more difficult environment where less theoretical guidance is available on the properties of the data.

\subsection{Macroeconomic forecasting using the FRED-MD dataset} \label{sec:FRED}

We consider forecasting eight US macroeconomic variables from the FRED-MD dataset at 1, 6 and 12 months forecast horizons. We first focus on the strategy discussed in Section \ref{sec:trans} where we first transform all series to I(0) before estimating the forecasting models. We illustrate the unit root testing methods, and show the empirical consequences of specification changes in the orders of integration. Next, we analyze the methods discussed in Section \ref{sec:coint}, and compare their forecast accuracy.

\subsubsection{Transformations to stationarity} \label{sec:OiI_FRED}

As the FRED-MD series have already been classified by \citet{McCrackenNg16}, we have a benchmark for our own classification using the unit root testing methodology discussed in Section \ref{sec:trans}. We consider the autoregressive wild bootstrap as described in Section \ref{sec:mboot} in combination with the union test in \eqref{eq:URunion}. We set the AWB parameter $\gamma$ equal to 0.85, which implies that over a year of serial dependence is captured by the bootstrap. Lag lengths in the ADF regressions are selected by the rescaled MAIC criterion of \citet{CPST15}, which is robust to heteroskedasticity. To account for multiple testing, we control the false discovery rate at 5\% using the bootstrap method of \citet{RSW08fdr} (labelled as `BFDR') and apply the sequential test procedure of \citet{Smeekes15} (labelled as `BSQT') with a significance level of 5\% and evenly spaced 0.05 quantiles such that $p_k = [0.05(k-1)]$ for $k=1, \ldots, 20$. We also perform the unit root tests on each series individually (labelled as `iADF') with a significance level of 5\%.

As some series in the FRED-MD are likely $I(2)$, we need to extend the methodology to detect these as well. We consider two ways to do so. First, we borrow information about the $I(2)$ series from the official FRED-MD classification, and take first differences of the series deemed to be $I(2)$. We then put these first differences together with the other series in levels and test for unit roots. This strategy ensures that the $I(2)$ series are classified at least as $I(1)$, and we only need to perform a single round of unit root testing. Our second approach is fully data-driven and follows a multivariate extension of the ``Pantula principle'' \citep{Pantula89}, where we first test for a unit root in the first difference of all series. The series for which the null cannot be rejected are classified as $I(2)$ and removed from the sample. The remaining series are then tested in levels and consequently classified as $I(1)$ or $I(0)$. In the results we append an acronym with a 1 if the first strategy is followed, and with a 2 if the second strategy is followed.\footnote{We take logarithmic transformations of the series before differencing when indicated by the official FRED-MD classification. Determining when a logarithmic transformation is appropriate is a daunting task for such a high-dimensional system as it seems difficult to automatize, especially as it cannot be seen separately from the determination of the order of integration \citep{FransesMcAleer98,KramerDavies02}. \citet{KKS17} propose a high-dimensional method to determine an appropriate transformation model, but it is not trivial how to combine their method with unit root testing. Therefore we apply the ``true'' transformations such that we can abstract from this issue.}

As a final method, we include a ``naive'' unit root testing approach that we believe is representative of casual unit root testing applied by many practitioners who, understandably, may not pay too much detailed attention to the unit root testing. In particular, we use the \texttt{adf.test} function from the popular R package `tseries' \citep{tseries}, and apply it with its default options, which implies performing individual ADF tests with a trend and setting a fixed lag length as a function of the sample size.\footnote{The lag length is set equal to $\lfloor(T-1)^{1/3}\rfloor$.} Our goal is not to discuss the merits of this particular unit root test procedure, but instead to highlight the consequences of casually using a ``standard'' unit root test procedure that does not address the issues described in Section \ref{sec:trans}.

\begin{figure}
    \centering
    \includegraphics[width=0.85\textwidth]{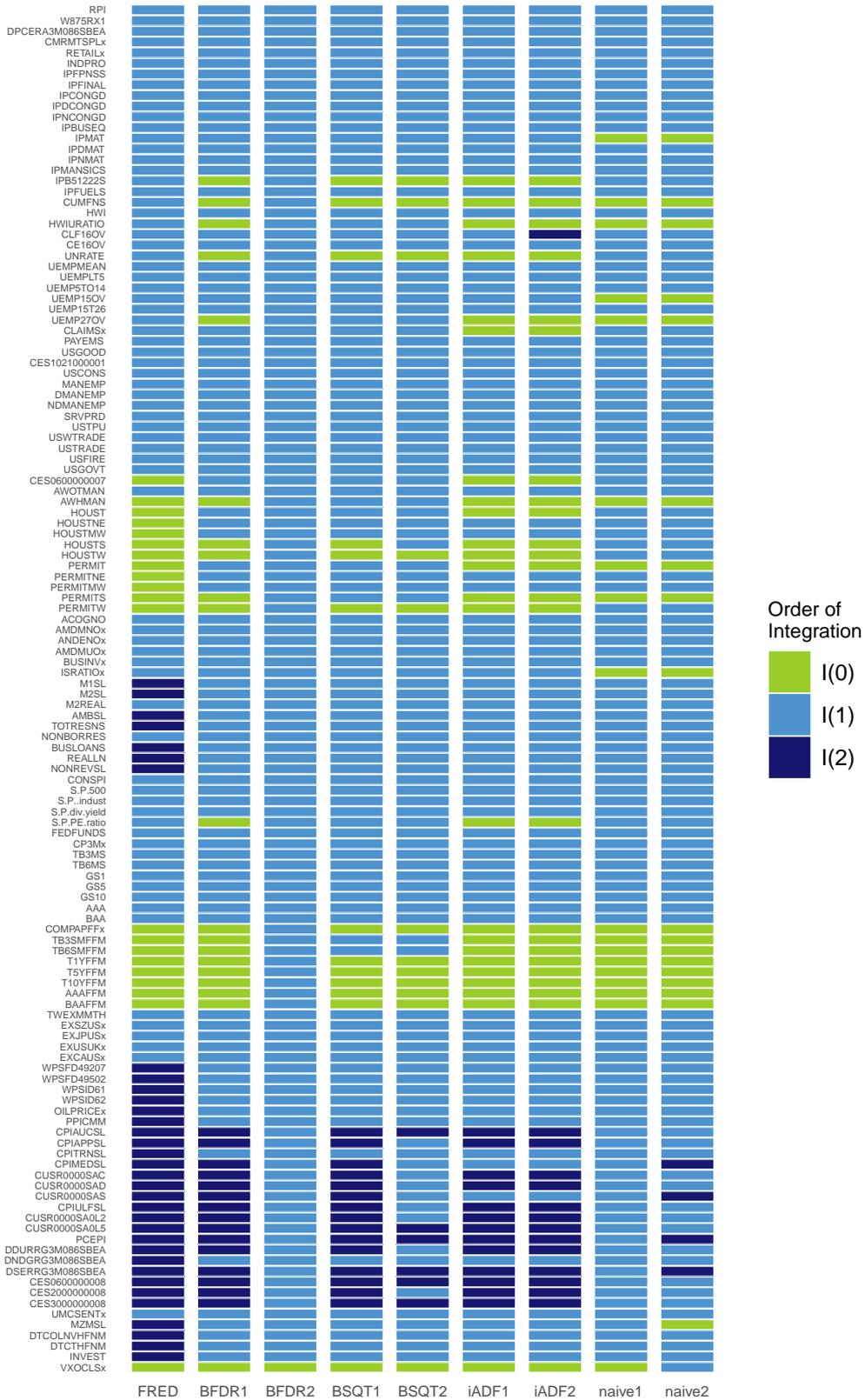}
    \caption{Classification of the FRED-MD dataset into I(0), I(1) and I(2) series.}
    \label{fig:OoI_FRED}
\end{figure}

Figure \ref{fig:OoI_FRED} presents the found orders of integration. Globally the classification appears to agree among the different methods, which is comforting, although some important differences can be noted. First, none of the data-driven methods finds as many $I(2)$ series as the FRED classification does. Indeed, this may not be such a surprising result, as it remains a debated issue among practitioners whether these series should be modelled as $I(1)$ or as $I(2)$, see for example the discussion in \citet{MSW06}.

Second, although most methods yield fairly similar classifications, the clear outlier is BFDR2, which finds all series but one to be $I(1)$. The FDR controlling algorithm may, by construction, be too conservative in the early stages of the algorithm when few rejections $R$ have been recorded, yet too liberal in the final stages upon finding many rejections. Indeed, when testing the first differences of all series for a unit root, the FRED classification tells us that for most of the series the null can be rejected. When the algorithm arrives at the $I(2)$ series, the unit root hypothesis will already have been rejected for many series. With $R$ being that large, the number of false rejections $F$ can be relatively large too without increasing the FDR too much. Hence, incorrectly rejecting the null for the $I(2)$ series will fall within the ``margins of error'' and thus lead to a complete rejection of all null hypotheses. In the second step the FDR algorithm then appears to get ``stuck'' in the early stages, resulting in only a single rejection. This risk of the method getting stuck early on was also observed by \citet{Smeekes15} and can be explained by the fact that early on in the step-down procedure, when $R$ is small, FDR is about as strict as FWE. It appears that in this case the inclusion of the $I(2)$ series in levels rather than differences is just enough to make the algorithm get stuck.

Third, even though iADF does not control for multiple testing, its results are fairly similar to BSQT and FDR1. It therefore appears explicitly controlling for multiple testing is not the most important in this application, and sensible unit root tests, even when applied individually, will give reasonable answers. On first glance even using the `naive' strategy appears not be very harmful. However, upon more careful inspection of the results, we can see that it does differ from the other methods. In particular, almost no $I(2)$ series are detected by this strategy, and given that there is no reason to prefer it over the other methods, we recommend against its use.

\subsubsection{Forecast comparison after transformations}\label{sec:forecast_stationary_FRED}
While determining an appropriate order of integration may be of interest in itself, our goal here is to evaluate its impact on forecast accuracy. As such, we next evaluate if, and how, the chosen transformation impacts the actual forecast performance of the BFDR, BSQT and iADF methods, all in both strategies considered, in comparison with the official FRED classification.

We forecast eight macroeconomic series in the FRED-MD dataset using data from July 1972 to October 2018. The series of interest consist of four real series, namely real production income (RPI), total industrial production (INDPRO), real manufacturing and trade industries sales (CMRMTSPLx) and non-agricultural employees (PAYEMS), and four nominal series, being the producer index for finished good (WPSFD49207), consumer price index - total (CPIAUCSL), consumer price index - less food (CPIULFSL) and the PCE price deflator (PCEPI). Each series is forecast $h$ months ahead, where we consider the forecast horizons $h=1,6,12$. All models are estimated on a rolling window spanning ten years, i.e. containing 120 observations. Within each window, we regress every time series on a constant and linear trend and obtain the corresponding residuals. For the stationary methods, these residuals are transformed to stationarity according to the results of the unit root testing procedure. Each model is fitted to these transformed residuals, after which the $h$-step ahead forecast is constructed as an iterated one-step-ahead forecast, when possible, and transformed to levels, if needed. The final forecast is obtained by adding the level forecast of the transformed residuals to the forecast of the deterministic components. We briefly describe the implementation of each method below. 

We consider four methods here. The first method is a standard vector autoregressive (VAR) model, fit on the eight variables of interest. Considering only the eight series of interest, however, may result in a substantial loss of relevant information contained in the remaining variables in the complete data set. Therefore, we also consider a factor-augmented vector autoregressive model (FAVAR) in the spirit of \citet{BBE05}, which includes factors as proxies for this missing information. We extract four factors from the complete and transformed data set and fit two separate FAVAR models containing these four factors, in addition to either the four real or the four nominal series. Rather than focusing on the estimation of heavily parameterized full systems, one may attempt to reduce the dimensionality by considering single-equation models, as discussed in Section \ref{sec:SE}. Conditioning the variable of interest on the remaining variables in the data set, results in an autoregressive distributed lag model with $M=N(p+1)-1$ parameters. For large $N$, shrinkage may still be desirable. Therefore, we include a penalized autoregressive distributed lag model (PADL) in the comparison, which is based on the minimization of 
\begin{equation}\label{eq:PADL}
    \sum_{t=1}^T\left(y^h_t - \bpi^\prime \bw_t\right)^2 + \lambda\sum_{j=1}^{M}\omega_{\pi,j}^{k_\pi}\abs{\pi}_j,
\end{equation}
where
\begin{equation}\label{eq:yh}
    y^h_t = \begin{cases}
    y_{t+h}-y_t & \text{ if } y_t \sim I(1),\\
    y_{t+h}-y_t - \Delta y_t & \text{ if } y_t \sim I(2).
    \end{cases}
\end{equation}
Furthermore,  $\bw_t$ contains contemporaneous values of all transformed time series except $y_t$, and three lags of all transformed time series. The weights $\omega_{\pi,j}^{k_\pi}$ are as defined in Section \ref{sec:SE}. In essence, this can be seen as an implementation of SPECS with the build-in restriction that $\bdelta=\bm{0}$, thereby ignoring cointegration. Finally, the concept of using factors as proxies for missing information remains equally useful for single-equation models. Accordingly, we include a factor-augmented penalized autoregressive distributed model (FAPADL) which is a single-equation model derived from a FAVAR. We estimate eight factors on the complete data set, which are added to the eight variables of interest in a the single-equation model. This is then estimated in accordance to \eqref{eq:PADL}, with $\bw_t$ now containing contemporaneous values and three lags of the eight time series of interest and the eight factors. The PADL and FAPADL are variants of the adaptive lasso and we implement these in R based on the popular `glmnet' package \citep{Friedman2009}. The lag order for the VAR and FAVAR are chosen by the BIC criterion, with a maximum lag order of three.

Oar goal is not be exhaustive, but we believe these four methods cover a wide enough range of available high-dimensional forecast methods such that our results cannot be attributed to the choice of a particular forecasting method and instead genuinely reflect the effect of different transformations to stationarity. For the sake of space, we only report the results based on the FAVAR here for 1 month and 12 moths ahead forecasts, as these are representative for the full set of results (which are available upon request). Generally, we find the same patterns within each method as we observe for the FAVAR, though they may be more or less pronounced. Overall the FAVAR is the most accurate of the four methods considered, which is why we choose to focus on it.

We compare the methods through their relative Mean Squared Forecast Errors (MSFEs), where the AR model is taken as benchmark. To attach a measure of statistical significance to these MSFEs, we obtain 90\% Model Confidence Sets (MCS) of the best performing model. We obtain the MCS using the autoregressive wild bootstrap as in \citet{SmeekesWijler18}. For the full details on the MCS implementation we refer to that paper.

\begin{figure}
    \centering
    \includegraphics[width=\textwidth]{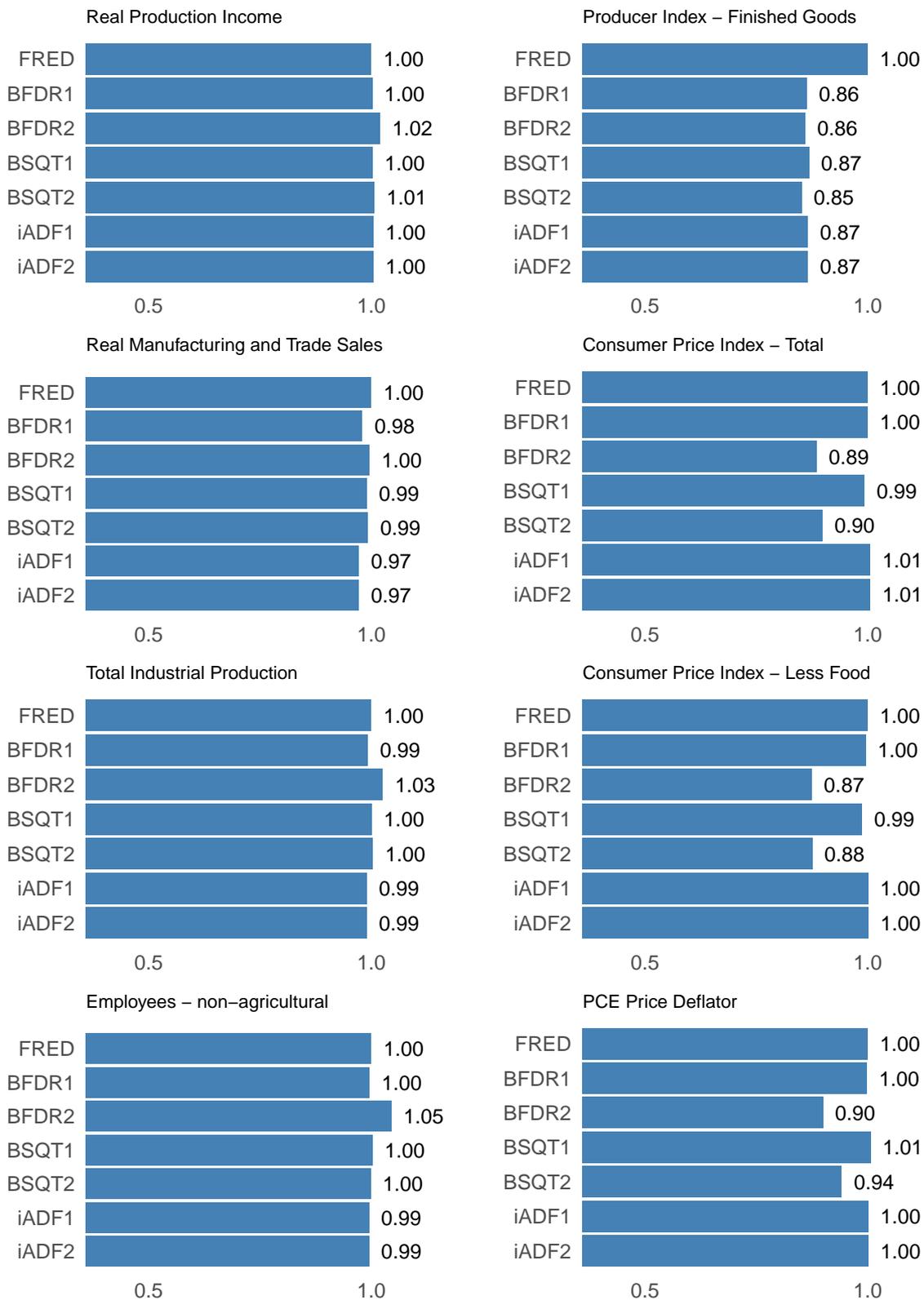}
    \caption{MCS (in blue) and relative MSFEs for 1-month horizon.}
    \label{fig:for1mF}
\end{figure}

\begin{figure}
    \centering
    \includegraphics[width=\textwidth]{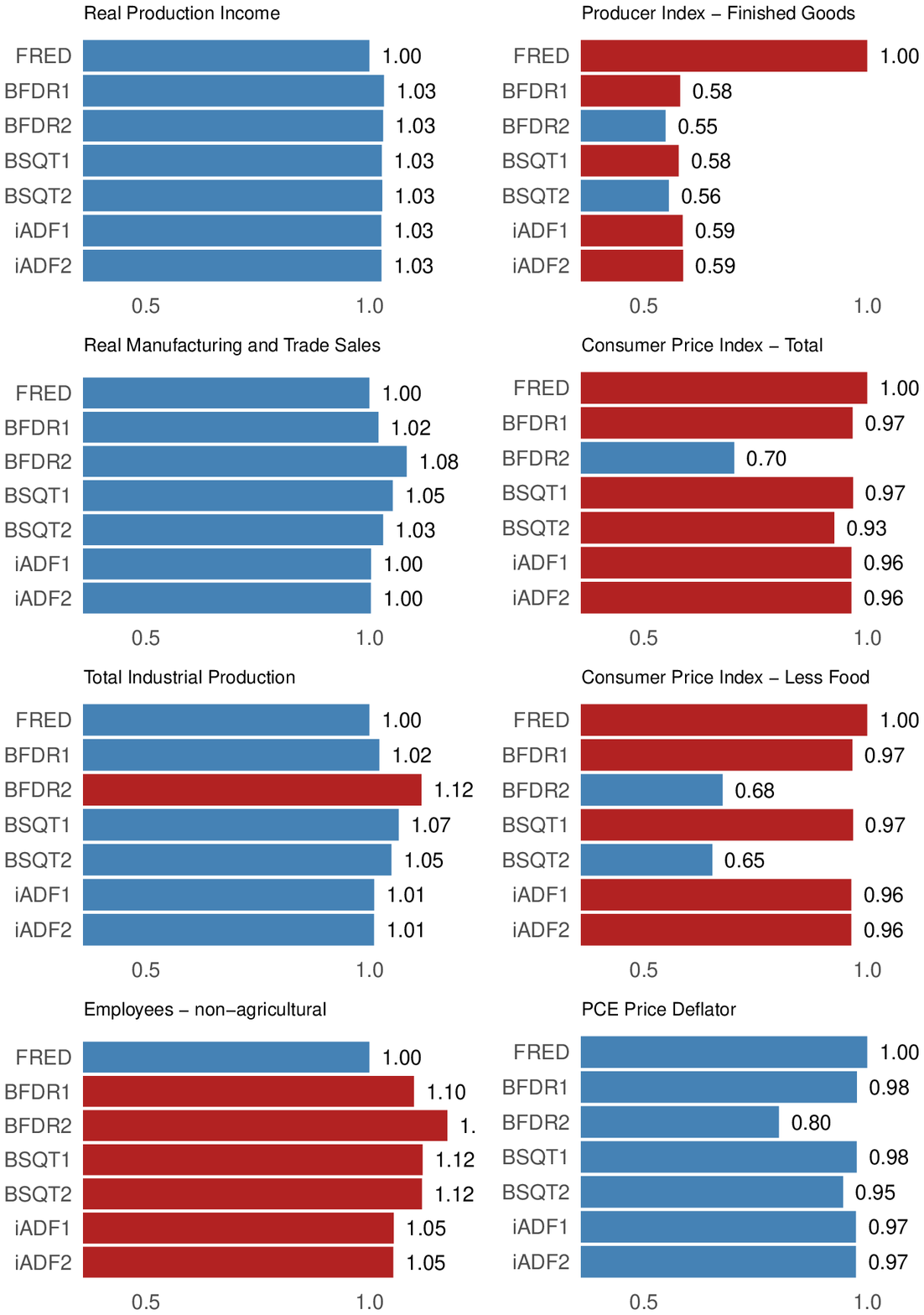}
    \caption{MCS (in blue) and relative MSFEs for 12-month horizon.}
    \label{fig:for12mF}
\end{figure}

The results are given in Figures \ref{fig:for1mF} and \ref{fig:for12mF}. For the one-month-ahead forecast the results are close for the different transformation methods, but for the twelve-months-ahead forecasts, we clearly see big differences for the nominal series. Inspection of the classifications in Figure \ref{fig:OoI_FRED} shows that the decisive factor is the classification of the dependent variable. For the three price series, the methods that classify these as $I(1)$ rather than $I(2)$ obtain substantial gains in forecast accuracy. Interestingly, the FRED classification finds these series to be $I(2)$, and thus deviating from the official classification can lead to substantial gains. These results are in line with the results of \citet{MSW06}, who also find that modelling price series as $I(1)$ rather than $I(2)$ results in better forecast accuracy.

As the outlying BFDR2 classification also classifies these series as I(1), this ``lucky shot'' eclipses any losses from the miss-classification of the other series. However, for the real series it can be observed that BFDR2 does indeed always perform somewhat worse than the other methods, although the MCS does not find it to be significant everywhere.

Concluding, miss-classification of the order of integration can have an effect on the performance of high-dimensional forecasting methods. However, unless the dependent variable is miss-classified, the high-dimensional nature of the data also ensures that this effect is smoothed out. On the other hand, correct classification of the dependent variable appears to be crucial, in particular regarding the classification as $I(1)$ versus $I(2)$.

\subsubsection{Forecast comparisons for cointegration methods} \label{sec:coint_FRED}

The forecast exercise for the methods that are able to take into account the cointegrating properties of the data proceeds along the same lines as in Section \ref{sec:forecast_stationary_FRED}. A noteworthy exception is that the time series that are considered I(1) in the FRED-MD classification are now kept in levels, whereas those that are considered as I(2) are differenced once. The methods included in the comparison are: (i) the factor error correction model (FECM) by \citet{BMM14,BMM16,BMM17}, (ii) the non-stationary dynamic factor model (N-DFM) by \citet{BLL17,BLL18}, (iii) the maximum-likelihood procedure (ML) by \citet{Johansen95}, (iv) the QR-decomposed VECM (QR-VECM) by \citet{LiangSchienle19}, (v) the penalized maximum-likelihood (PML) by \citet{WilmsCroux16}, (vi) the single-equation penalized error correction selector (SPECS) by \citet{SmeekesWijler18specs} and (vii) a factor-augmented SPECS (FASPECS). The latter method is simply the single-equation model derived from the FECM, based on the same principles as the FAPADL from the previous section. It is worth noting that the majority of these non-stationary methods have natural counterparts in the stationary world; the ML procedure compares directly to the VAR model, FECM compares to FAVAR, and SPECS and FA-SPECS to PADL and FAPADL, respectively. Finally, all methods are compared against an AR model fit on the dependent variable, the latter being transformed according to the original FRED codes.

We briefly discuss some additional implementation choices for the non-stationary methods. For all procedures that require an estimate of the cointegrating rank, we use the information criteria proposed by \citet{ChengPhillips09}. The only exception is the PML method, for which the cointegrating rank is determined by the procedure advocated in \citet{WilmsCroux16}. Similar to \citet{BMM14}, we do not rely on information criteria to select the number of factors, but rather fix the number of factors in the implementation of the FECM and N-DFM methods to four.\footnote{In untabulated results, we find that the forecast performance does not improve when the number of factors is selected by the information criteria by \citet{Bai04}. Neither does the addition of a stationary factor computed from the estimated idiosyncratic component, in the spirit of \citet{BMM14}. Both strategies are therefore omitted from the analysis.} In the N-DFM approach, we model the idiosyncratic components of the target variables as simple AR models. The ML procedure estimates a VECM system on the eight variables of interest. In congruence with the implementation of the stationary methods, the lag order for FECM, N-DFM and ML is chosen by the BIC criterion, with a maximum lag order of three. The QR-VECM and PML methods are estimated on a dataset containing the eight series of interest and an additional 17 variables, informally selected based on their unique information within each economic category. Details are provided in Table \ref{tb:varsFRED}. We incorporate only a single lag in the QR-VECM implementation, necessitated by the requirement of initial OLS estimates. SPECS estimates the model
\begin{equation*}
    y^h_t = \bdelta^\prime \bz_{t-1} + \bpi^\prime \bw_t + \epsilon_{y,t},
\end{equation*}
where $y^h_t$ is defined in \eqref{eq:yh}, with the order of integration based on the original FRED codes. Note that the variables included in $\bz_t$ are either the complete set of 124 time series or the eight time series of interest plus an additional eight estimated factors, depending on whether the implementation concerns SPECS or FA-SPECS, respectively. Finally, all parameters that regulate the degree of shrinkage are chosen by time series cross-validation, proposed by \citet{Hyndman2018} and discussed in a context similar to the current analysis in \citet[][p. 411]{SmeekesWijler18}.

\begin{table}[t]
\centering
\caption{Overview of the variables included for QR-VECM and PML.} \label{tb:varsFRED}
\begin{tabular}{cll}
\hline 
 & \textbf{FRED code} & \textbf{description}\tabularnewline
\hline 
\hline 
\multirow{4}{*}{\begin{turn}{90}
Real
\end{turn}} & RPI & Real Personal Income\tabularnewline
 & CMRMTSPLx & Real Manufacturing and Trades Industries Sale\tabularnewline
 & INDPRO & IP Index\tabularnewline
 & PAYEMS & All Employees: Total nonfarm\tabularnewline
\hline 
\multirow{4}{*}{\begin{turn}{90}
Nominal
\end{turn}} & WPSFD49207 & PPI: Finished Goods\tabularnewline
 & CPIAUCSL & CPI : All Items\tabularnewline
 & CPIULFSL & CPI : All Items Less Food\tabularnewline
 & PCEPI & Personal Cons. Expend.: Chain Index\tabularnewline
\hline 
\multirow{17}{*}{\begin{turn}{90}
Additional
\end{turn}} & CUMFNS & Capacity Utilization: Manufacturing\tabularnewline
 & HWI & Help-Wanted Index for United States\tabularnewline
 & UNRATE & Civilian Unemployment Rate\tabularnewline
 & UEMPMEAN & Average Duration of Unemployment (Weeks)\tabularnewline
 & HOUST & Housing Starts: Total New Privately Owned\tabularnewline
 & PERMIT & New Private Housing Permits (SAAR)\tabularnewline
 & BUSINVx & Total Business Inventories\tabularnewline
 & M1SL & M1 Money Stock\tabularnewline
 & M2SL & M2 Money Stock\tabularnewline
 & FEDFUNDS & Effective Federal Funds Rate\tabularnewline
 & TB3MS & 3-Month Treasury Bill\tabularnewline
 & GS5 & 5-Year Treasury Rate\tabularnewline
 & GS10 & 10-Year Treasury Rate\tabularnewline
 & EXJPUSx & Japan / U.S. Foreign Exchange Rate\tabularnewline
 & EXUSUKx & U.S. / U.K. Foreign Exchange Rate\tabularnewline
 & EXCAUSx & Canada / U.S. Foreign Exchange Rate\tabularnewline
 & S.P.500 & S\&P Common Stock Price Index: Composite\tabularnewline
\hline 
\end{tabular}
\end{table}

\begin{figure}
    \centering
    \includegraphics[width=\textwidth]{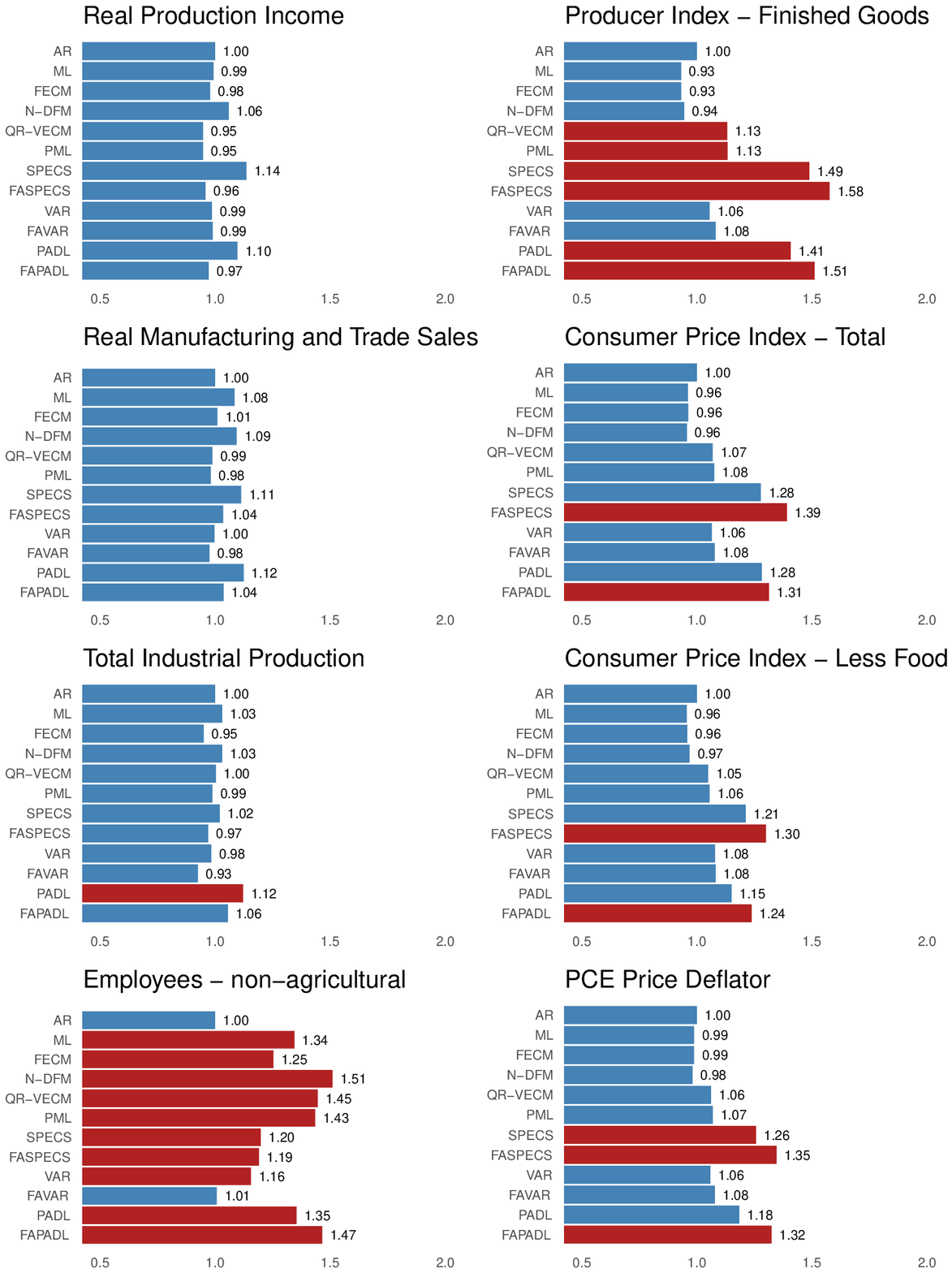}
    \caption{MCS (in blue) and relative MSFEs for 1-month horizon.}
    \label{fig:for1m}
\end{figure}

\begin{figure}
    \centering
    \includegraphics[width=\textwidth]{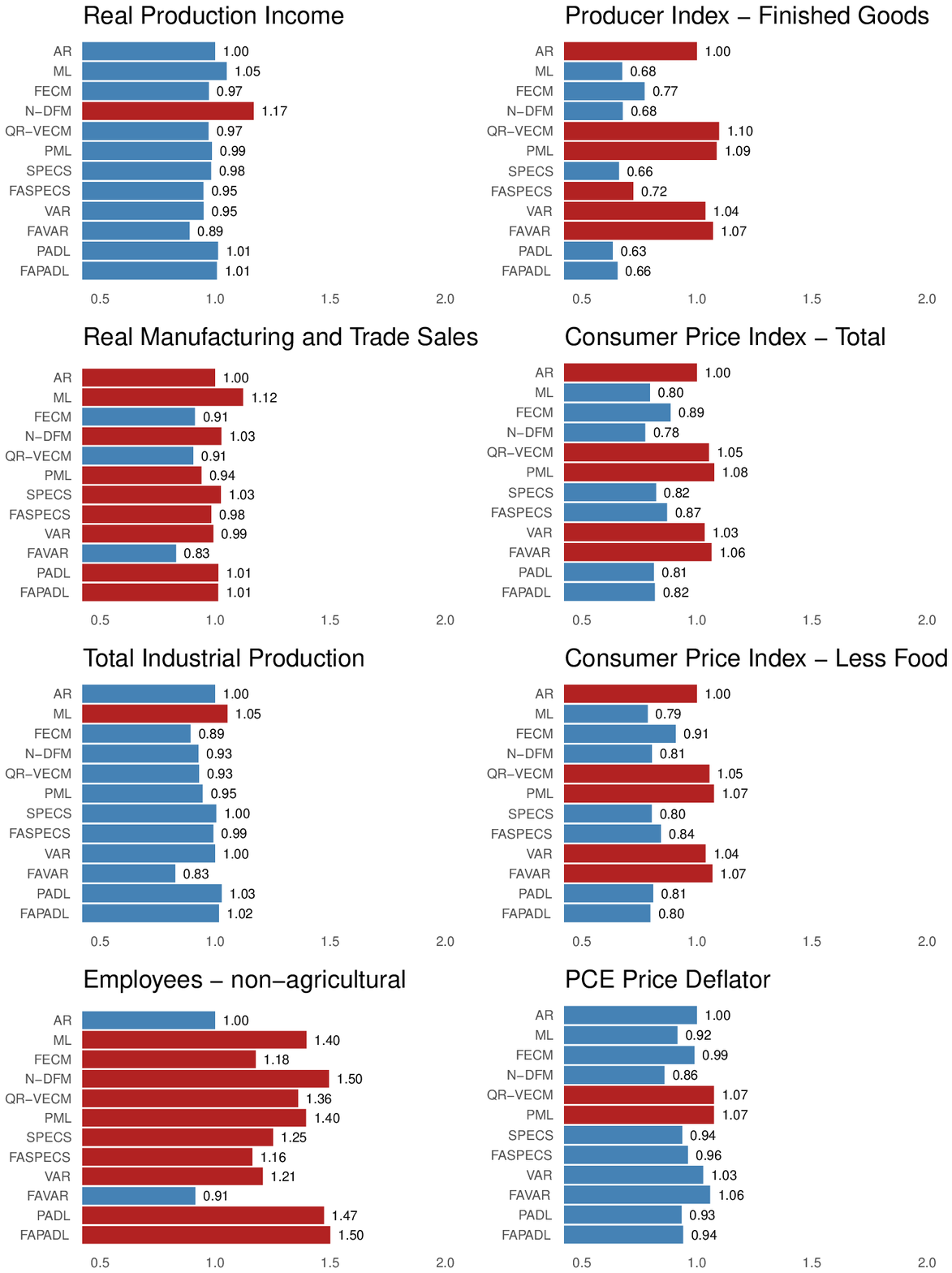}
    \caption{MCS (in blue) and relative MSFEs for 6-month horizon.}
    \label{fig:for6m}
\end{figure}

\begin{figure}
    \centering
    \includegraphics[width=\textwidth]{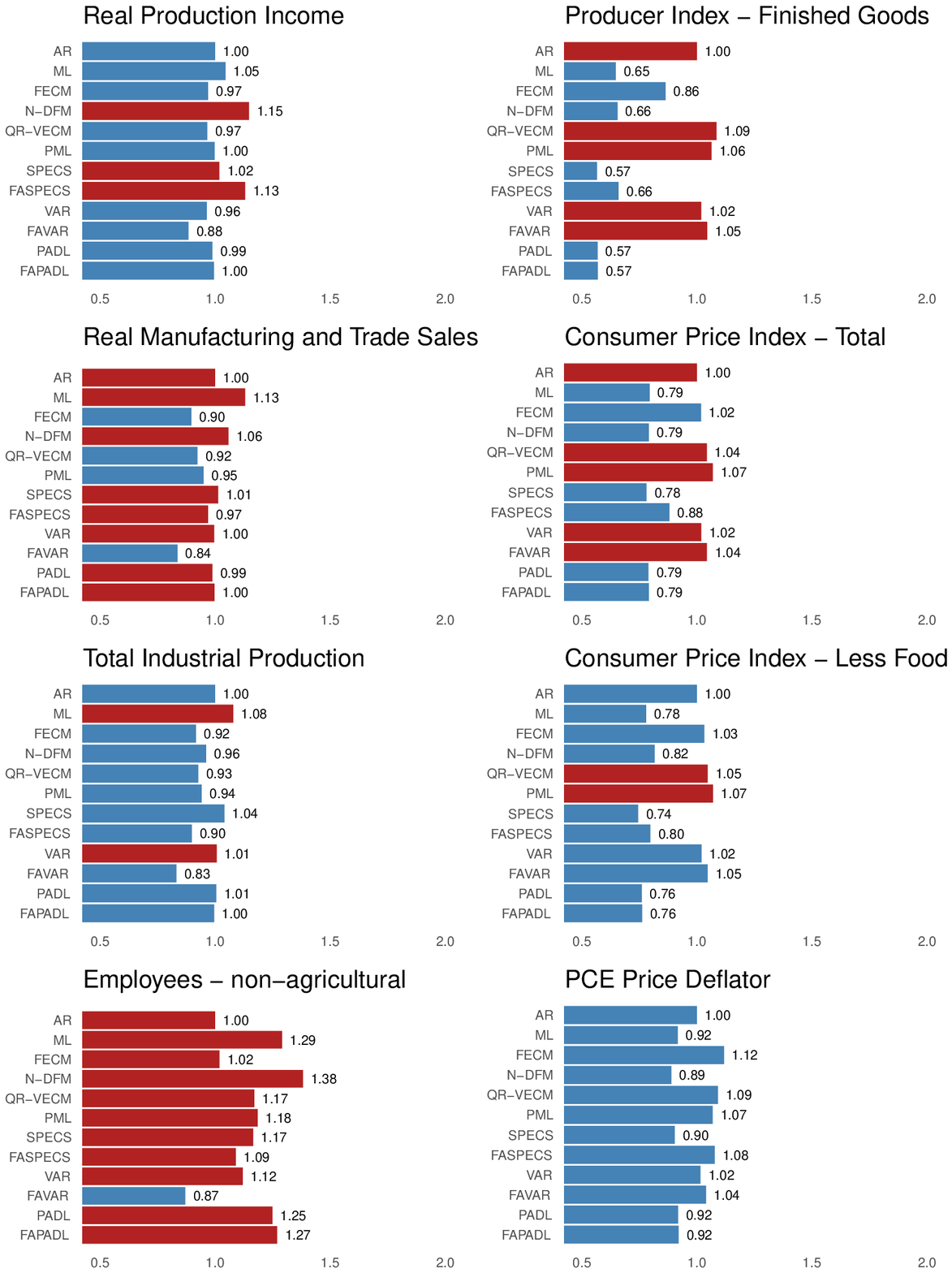}
    \caption{MCS (in blue) and relative MSFEs for 12-month horizon.}
    \label{fig:for12m}
\end{figure}

Results are given in Figure \ref{fig:for1m}-\ref{fig:for12m}. Considering first the 1-month ahead predictions, we observe similar forecasting performance on the first three real series (RPI,CMRMTSPLx, INDPRO) with almost none of the methods being excluded from the 90\% model confidence set. The employment forecasts of the AR benchmark and the FAVAR approach are considered superior to those of the other methods. On the four nominal series, the sparse high-dimensional methods display relatively poor performance, regardless of whether they take into account potential cointegration in the data. Overall, no clear distinction is visible between the non-stationary and stationary methods, although this may not come as a surprise given the short forecast horizon. As usual, the AR benchmark appears hard to beat and is not excluded from any of the model confidence sets here.

The forecast comparisons for longer forecast horizons display stronger differentiation across methods. Our findings are qualitatively similar for the 6-month and 12-month horizons, and, for the sake of brevity, we comment here on the 12-month horizon only. The results for the first three real series again do not portray a preference for taking into account cointegration versus transforming the data. Comparing ML and VAR to FECM and FAVAR, incorporating information across the whole data set seems to positively affect forecast performance, a finding that is additionally confirmed by the favourable performance of the penalized VECM estimators. The FAVAR substantially outperforms on the employment series, being the only method included in the model confidence set. On the nominal series, the single-equation methods perform well, again not showing any gain or loss in predictive power by accounting for cointegration. The ML and N-DFM procedure methods show favourable forecast accuracy as well, whereas the two penalized VECM estimators appear inferior on the nominal series. The AR benchmark is excluded for four out of eight series.

In summary, the comparative performance is strongly dependent on the choice of dependent variable and forecast horizon. For short forecast horizons, hardly any statistically significant differences in forecast accuracy are observed. However, for longer horizons the differences are more pronounced, with factor-augmented or penalized full system estimators performing well on the real series, the FAVAR strongly outperforming on the employment series, and the single-equation methods appearing superior on the nominal series. The findings do not provide conclusive evidence whether cointegration matters for forecasting.

\subsection{Unemployment Nowcasting with Google Trends} \label{sec:URGT}
In this section we revisit the nowcasting application of \citet{SmeekesWijler18specs}, who consider nowcasting unemployment using Google Trends data. One of the advantages of modern Big Datasets is that information obtained from internet activity is often available on very short notice, and can be used to supplement official statistics produced by statistical offices. For instance, internet searches about unemployment-related issues may contain information about people being or becoming unemployed, and could be used to obtain unemployment estimates before statistical offices are able to produce official unemployment statistics. 

Google records weekly and monthly data on the popularity of specific search terms through its publicly available Google Trends service,\footnote{\url{https://trends.google.com/trends}} with data being available only days after a period ends. On the other hand, national statistical offices need weeks to process surveys and produce official unemployment figures for the preceding month. As such, Google Trends data on unemployment-related queries would appear to have the potential to produce timely nowcasts of the latest unemployment figures.

Indeed, \citet{SPSB19} propose a dynamic factor model within a state space context to combine survey data with Google Trends data to produce more timely official umemployment statistics. They illustrate their method using a dataset of about one hundred unemployment-related queries in the Netherlands obtained from Google Trends. \citet{SmeekesWijler18specs} consider a similar setup with the same Google Trends data, but consider the conceptually simpler setup where the dependent variable to be nowcasted is the official published unemployment by Statistics Netherlands.\footnote{Additionally, this means the application does not require the use of the private survey data and is based on publicly available data only.} Moreover, they exclusively focus on penalized regression methods. In this section we revisit their application in the context of the methods discussed here. For full details on the dataset, which is available on the authors' websites, we refer to \citet{SmeekesWijler18specs}.

\subsubsection{Transformations to stationarity} \label{sec:OiI_URGT}

As for the FRED-MD dataset, we first consider the different ways to classify the series into I(0), I(1) and I(2) series. However, unlike for the FRED data, here we don't have a pre-set classification available, and therefore unit root testing is a necessity before continuing the analysis. Moreover, as the dataset could easily be extended to an arbitrarily high dimension by simply adding other relevant queries, an automated fully-data driven method is required. 

This lack of a known classification also means that our first strategy as used in Section \ref{sec:OiI_FRED} has to be adapted, as we cannot differ I(2) series a priori. In particular, for our first strategy we assume that the series can be at most I(1), and hence we perform only a single unit root test on the levels of all series. Our second strategy is again the Pantula principle as in Section \ref{sec:OiI_FRED}. Within each strategy we consider the same four tests as before.\footnote{As Google reports the search frequencies in relative terms (both to the past and other searches), we do not take logs anywhere.}

\begin{figure}
    \centering
    \includegraphics[width=0.85\textwidth]{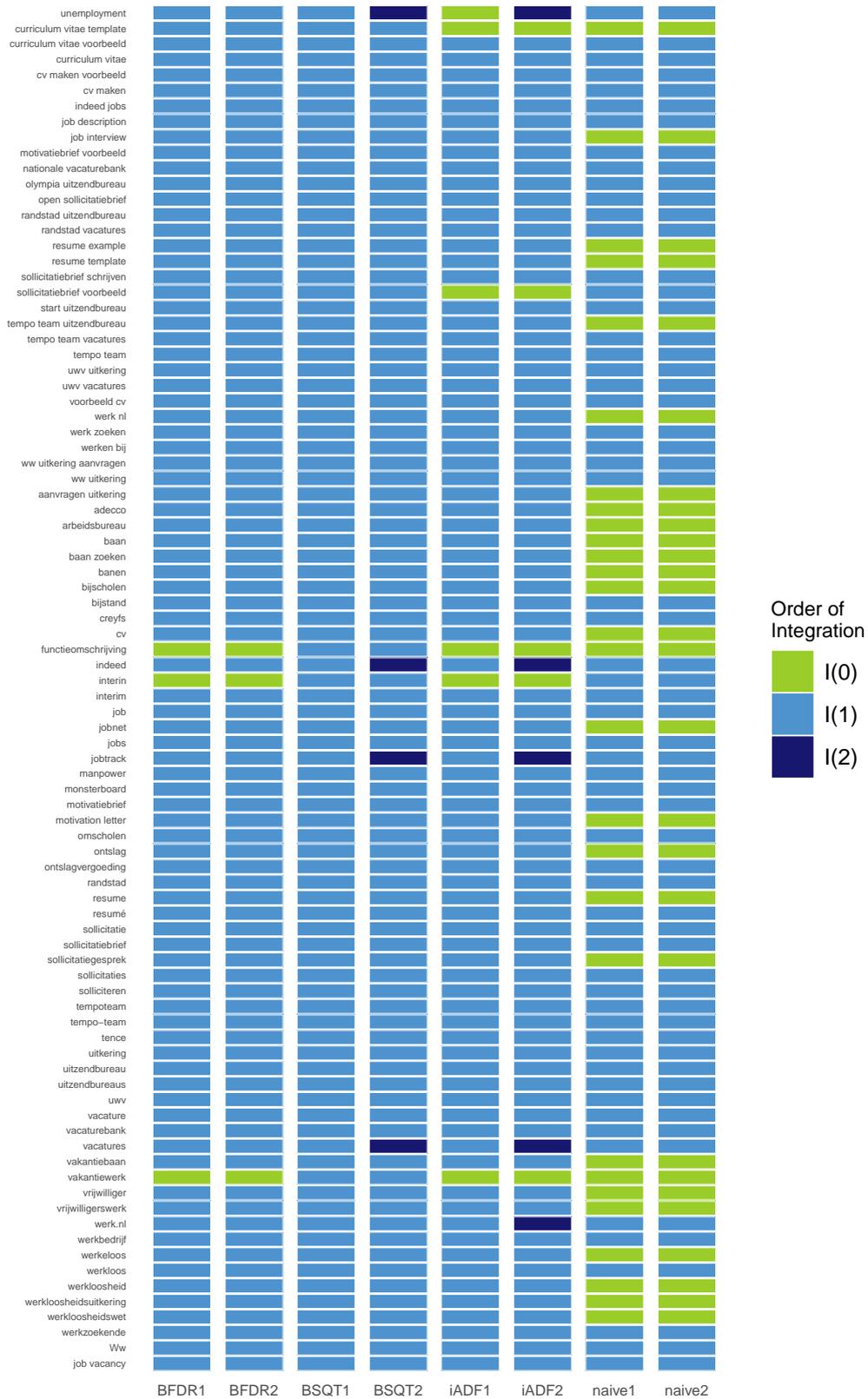}
    \caption{Classification of unemployment dataset into I(0), I(1) and I(2) series.}
    \label{fig:OoI_URGT}
\end{figure}

The classification results are given in Figure \ref{fig:OoI_URGT}. Generally they provide strong evidence that nearly all series are I(1), with most methods only finding very few I(0) and I(2) series. Interestingly, one of the few series that the methods disagree about is the unemployment series, which receives all three possible classifications. From our previous results we may expect this series, our dependent variable, to be the major determinant of forecast accuracy. Aside from this result, the most striking result is the performance of the naive tests, that find many more I(0) variables than the other methods. One possible explanation for this result may be the nature of the Google Trends data, that can exhibit large changes in volatility. As standard unit root tests are not robust to such changes, 
a naive strategy might seriously be affected, as appears to be the case here.

\subsubsection{Forecast comparison} \label{sec:forecast_URGT}
We now compare the nowcasting performance of the high-dimensional methods. Given our focus on forecasting the present, that is $h=0$, for a single variable, there is little benefit in considering the system estimators we used before. Therefore we only consider the subset of single-equation models that allow for nowcasting. Specifically, we include SPECS as described in Section \ref{sec:SE} as well as its modification FA-SPECS described in Section \ref{sec:coint_FRED} as methods that explicitly account for unit roots and cointegration. Furthermore, we include PADL and FAPADL as described in Section \ref{sec:forecast_stationary_FRED}. For all methods, the modification for nowcasting is done by setting $h=0$, where we implicitly assume that at time $t$ the values for the explanatory variables are available, but that for unemployment is not. This corresponds to the real-life situation. 

For SPECS we model unemployment as (at most) $I(1)$, given that this is its predominant classification in Figure \ref{fig:OoI_URGT}. Additionally, we include all regressors in levels, thereby implicitly assuming these are at most $I(1)$ as well, which is again justified by the preceding unit root tests. For PADL and FAPADL we transform the series to stationarity according to the obtained classifications. Again we consider an AR model as benchmark, while all other implementational details are the same as in Section \ref{sec:FRED}.

Our dataset covers monthly data from January 2004 until December 2017 for unemployment obtained from Statistics Netherlands, and 87 Google Trends series. We estimate the models on a rolling window of 100 observations each, leaving 64 time periods for obtaining nowcasts. We compare the nowcast accuracy through relative Mean Squared Nowcast Error (MSNE), with the AR model as benchmark, and obtain 90\% Model Confidence Sets containing the best models in the same way as in Section \ref{sec:FRED}.

\begin{figure}[t]
    \centering
    \includegraphics[width=0.5\textwidth]{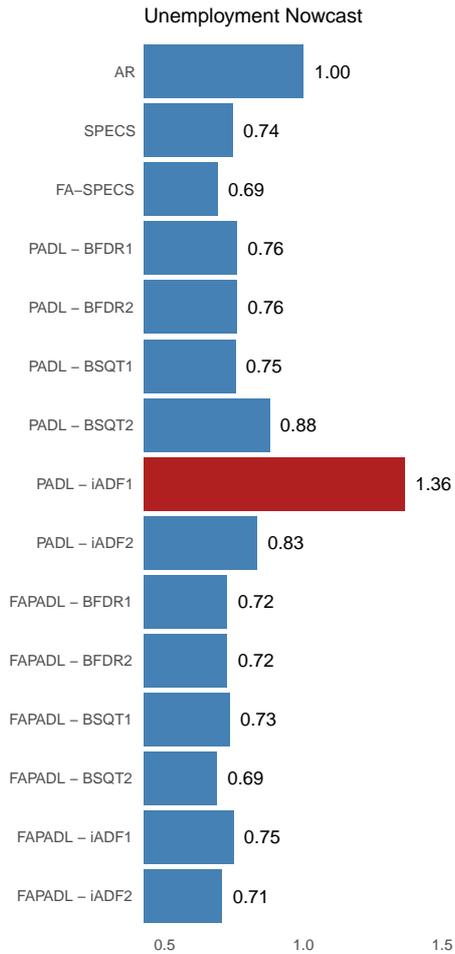}
    \caption{MCS (in blue) and relative MSNEs for unemployment nowcast.}
    \label{fig:now}
\end{figure}

Figure \ref{fig:now} presents the results. We see that, with the exception of the PADL - iADF1 method, all methods outperform the AR benchmark, although the 90\% MCS does not find the differences to be significant. Factor augmentation generally leads to slightly more accurate forecasts than the full penalization approaches, but differences are marginal. Interestingly, the classification of unemployment appears to only have a minor effect on the accuracy, with $I(0)$, $I(1)$ and $I(2)$ classifications all performing similarly. This does not necessarily contradict the results in Section \ref{sec:FRED}, as differences were only pronounced there for longer forecast horizons, whereas the forecast horizon here is immediate. Finally, we observe that the SPECS methods are always at least as accurate as their counterparts that do not take cointegration into account. It therefore seems to pay off to allow for cointegration, even though differences are again marginal.

\section{Conclusion} \label{sec:conc}
We investigated how the potential presence of unit roots and cointegration impacts macroeconomic forecasting in the presence of Big Data. We considered both the strategies of transforming all data to stationarity, and of explicitly modelling any unit roots and cointegrating relationships.

The strategy of transforming to stationarity is commonly thought of as allowing one to bypass the unit root issue. However, this strategy is not innocuous as often thought, as it still relies on a correct classification of the orders of integration of all series. Given that this needs to be done for a large number of series, there is a lot of room for errors, and naive unit root testing is not advised. We discussed potential pitfalls for this classification, and evaluated methods designed to deal with issues of poor size and power of unit root tests, as well as controlling appropriate error rates in multiple testing.

Next we considered modelling unit roots and cointegration directly in a high-dimensional framework. We reviewed methods approaching the problem from two different philosophies, namely that of factor models and that of penalized regression. Within these philosophies we also highlighted differences among the proposed methods both in terms of underlying assumptions and implementation issues.

We illustrated these methods in two empirical applications; the first considered forecasting macroeconomic variables using the well-established FRED-MD dataset, while the second considered nowcasting unemployment using Google Trends data. Both applications showed that transforming to stationarity requires careful considerations of the methods used. While the specific method used for accounting for multiple testing generally only led to marginal differences, a correct classification of the variable to be forecasted is critically important. We therefore recommend paying specific attention to these variables. Moreover, as occasionally some methods can deliver strange results, in general it is advisable to perform the classification using multiple approaches, to ensure that the classification found is credible.

The applications also demonstrated that there is no general way to model cointegration that is clearly superior. Indeed, the results do not show a clear conclusion on whether cointegration should be taken into account. This result, perhaps unsurprisingly, mirrors the literature on low-dimensional time series. It therefore remains up to the practitioner to decide for their specific application if, and if yes how, cointegration should be modelled for forecasting purposes. Overall, the methods we consider provide reliable tools to do so, should the practitioner wish to do so.

Concluding, several reliable tools are available for dealing with unit roots and cointegration in a high-dimensional forecasting setting. However, there is no panacea; a single best approach that is applicable in all settings does not exist. Instead, dealing with unit roots and cointegration in practice requires careful consideration and investigation which methods are most applicable in a given particular application. We also note that the field is rapidly developing, and major innovations are still to be expected in the near future. For instance, interval or density forecasting in high-dimensional systems with unit roots remains an entirely open issue. As high-dimensional inference is already complicated by issues such as post-selection bias, extending this to the unit root setting is very challenging indeed. Such tools however will be indispensable for the macroeconomic practitioner, and therefore constitute an exciting avenue for future research.

\singlespacing

\end{document}